\documentstyle[preprint,pre,eqsecnum,aps,amssymb,epsfig]{revtex}

\begin{document}

\tightenlines
\draft
\title{Quantitative study of laterally inhomogeneous wetting films}
\author{C. Bauer and S. Dietrich}
\address{Fachbereich Physik, Bergische Universit\"at Wuppertal,\\
D-42097 Wuppertal, Germany}

\maketitle

\begin{abstract}
Based on a microscopic density functional theory we calculate the
internal structure of the three-phase contact line between liquid,
vapor, and a confining wall as well as the
morphology of liquid wetting films on a substrate exhibiting a
chemical step. We present a refined numerical analysis 
of the nonlocal density functional which describes the
interface morphologies and the corresponding line
tensions. These results are compared with those predicted by a
more simple phenomenological interface displacement model. 
Except for the case that the interface exhibits large
curvatures, we find that the interface displacement model provides a
quantitatively reliable description of the interfacial structures.
\end{abstract}

\pacs{68.45.Gd,68.10.-m,82.65.Dp}

\section{Introduction}
\label{s:introduction}

The formation of a homogeneous wetting film of phase $\beta$ with
thickness $l_0$ at the planar $\alpha$-$\gamma$ interface between two
coexisting phases $\alpha$ and $\gamma$ is governed by the interplay
of the surface tensions $\sigma_{\alpha\beta}$, $\sigma_{\beta\gamma}$,
and $\sigma_{\alpha\gamma}$
\cite{degennesreview,sullivantelodagama,sdreview,schick}. In the
standard case $\beta$ and $\gamma$ are the liquid and vapor phases,
respectively, of a simple fluid and $\alpha$ represents a substrate
acting as an inert spectator phase. The translational invariance of
these systems in the lateral directions can be broken either by
opposing boundary conditions or by geometric or chemical
heterogeneities within the confining medium $\alpha$. Whereas the
first mechanism can always be applied, irrespective of the nature of
the phase $\alpha$, the latter two mechanisms require a solid phase
$\alpha$ which can permanently sustain well-defined lateral
structures.

As a prerequisite for the possibility to impose opposing boundary
conditions in lateral directions the thermodynamic state of the system
has to be chosen such that the two phases $\beta$ and $\gamma$ are in
thermal equilibrium. In that case the system can be arranged such that
on its left end the substrate $\alpha$ is exposed to the vapor phase
in the bulk whereas on its right end the substrate is exposed to the
coexisting liquid phase in the bulk. This arrangement leads to the
formation of a liquid-vapor interface intersecting the substrate at a
three-phase contact line with the contact angle $\theta =
\arccos((\sigma_{wg}-\sigma_{wl})/\sigma_{lg})$ (see
Fig.~\ref{f:cl_system}). The free energy of this configuration
decomposes into the volume contributions corresponding to the gas and
liquid phases, respectively, the surface tensions $\sigma_{wg}$,
$\sigma_{wl}$, and $\sigma_{lg}$ associated with the corresponding
half-planes of contact between these phases, and the line tension
$\tau$ associated with the presence of the three-phase contact line
\cite{degennesreview,gibbs,rowlinsonwidom}.

There have been numerous theoretical
(see, e.g., Refs.~\cite{berry,pethica,bennerscrivendavis,tarazonanavascues,toshevetal,varearobledo,szleiferwidom,bresmequirke,dobbs})
and experimental (for a review see, e.g., Ref.~\cite{drelich})
efforts to determine the sign, magnitude, and temperature dependence
of line tensions (see Refs.~\cite{getta} and \cite{sdsicily} for a
recent summary of an extended list of additional references). In
particular the singular behavior of the line tension upon approaching
wetting transitions, i.e., for $\theta\to 0$ has been examined by
using simple phenomenological interface displacement or
square-gradient models
\cite{dobbs,indekeu1992,indekeu1994,indekeudobbs,indekeurobledo}.
Since the long-ranged, i.e., power-law decay of the dispersion forces
acting in fluids is known to invalidate a gradient expansion for
effective interface Hamiltonians
\cite{napdiet1,napdiet2,napdiet3,napdiet4}, in Ref.~\cite{getta} the
actual nonlocal interface Hamiltonian, as it is obtained from a
microscopic nonlocal density functional theory, has been used to study
the influence of dispersion forces on the line tension and on the
intrinsic structure of the three-phase contact line. This study led to
significant quantitative differences in comparison with the more
simple interface displacement model.

Similar differences between the predictions of the nonlocal theory and
its local approximation appeared \cite{koch} in the analysis of the
morphology of a wetting film covering a planar substrate which
consists of two adjacent halves composed of different materials and
thus represents a chemical step (see
Fig.~\ref{f:step_system}). According to the results reported in
Ref.~\cite{koch} the nonlocal theory predicts a much broader
transition region, within which the local film thickness $l(x)$
switches as function of the lateral coordinate $x$ between its
asymptotic values $l_{\pm} = l(x\to\pm\infty)$, 
than the interface displacement model does.

The local interface displacement and square gradient theories relish
popularity and are convenient for the description of thin fluid films on
chemically or geometrically structured substrates (see, e.g.,
Refs.~\cite{colevittoratos,chengcole,kardarindekeu,robbinsetal,kochwedge,kagan,andersenjensen,swainparry,rasconparry,boulter,stellasartori,swainlipowsky}).
In views of the aforementioned known limitations of the applicability
of square gradient theories for systems governed by dispersion forces
\cite{napdiet1,napdiet2,napdiet3,napdiet4} the
objective of the present study is to pinpoint the reason for the
reported large quantitative differences following from the local and
the nonlocal approach. The common expectation is that the interface
displacement model and the square gradient
theory, although they are only approximations to the full nonlocal
theory, turn out to yield nonetheless reliable results for most of
the cases. The studies of the three-phase contact line
(Sec.~\ref{s:contactline}) and of the adsorption on a chemical step
(Sec.~\ref{s:chemicalstep}) serve as testing grounds for the
comparison between the local and the nonlocal theories. Our results
are summarized in Sec.~\ref{s:summary}.

\section{Three-phase contact line}
\label{s:contactline}

As stated in the Introduction we consider a simple fluid in a grand
canonical ensemble described by the chemical potential $\mu$ and the
temperature $T$. $\mu=\mu_0(T)$ is chosen such that in the bulk the
fluid is at liquid-vapor coexistence. This allows one to maintain the
configuration shown in Fig.~\ref{f:cl_system} which is characterized
by the number density $\rho(x,z)$ of the fluid particles. It
interpolates smoothly between $\rho(x\to-\infty,z)=\rho_-(z)$, which
corresponds to the wall-vapor interfacial profile, and
$\rho(x\to\infty,z)=\rho_+(z)$, which describes the wall-liquid
interfacial structure.
The local position $l(x)$ of the liquid-vapor interface can be
determined, e.g., as the isodensity contour line
$\rho(x,z=l(x))=\frac{1}{2}(\rho_l+\rho_g)$, where $\rho_l$ and
$\rho_g$ are the bulk densities in the liquid and vapor phase,
respectively. $l(x)$ asymptotically approaches the finite value
$l_0(T)$ for $x\to-\infty$, which corresponds to the equilibrium
wetting film thickness at the wall-vapor interface, and diverges 
linearly for $x\to\infty$ with a slope given by the contact angle (see
Fig.~\ref{f:cl_system}).

\subsection{Density functional theory}

Density functional theory has turned out to be a suitable theoretical
description for spatially inhomogeneous fluids as considered here. We
apply a simple \cite{evans} but nonetheless successful version which captures the
essentials of wetting transitions \cite{sdreview}.
Its grand canonical free energy functional reads
\begin{eqnarray}\label{e:functional}
\Omega([\rho({\mathbf r})];T,\mu) & = & \int_{\Lambda} d^3r
f_{HS}(\rho({\mathbf r}),T) + \int_{\Lambda} d^3r
[V({\mathbf r})-\mu]\rho({\mathbf r}) \nonumber\\
& & + \frac{1}{2}\int_{\Lambda} d^3r \int_{\Lambda} d^3r'
\tilde{w}(|{\mathbf r}-{\mathbf r}'|)\rho({\mathbf r})\rho({\mathbf r}'). 
\end{eqnarray}
$\Lambda$ is the finite volume filled by fluid within the half space
$\Lambda_+=\{{\mathbf r}\in {\mathbb R}^3|z>0\}$. In the thermodynamic limit one
has $\Lambda\to\Lambda_+$. The other half space $\{{\mathbf r}\in
{\mathbb R}^3|z\leq 0\}$ is occupied by the substrate. The external 
potential $V({\mathbf r})$ describes the interaction of a fluid
particle with the substrate:
\begin{equation}\label{e:substpothom}
V({\mathbf r}) = V(z>0) = -\sum\limits_{i\geq3} \frac{u_i}{z^i}.
\end{equation}
Approximately $V(z)$ can be thought of as a laterally averaged pairwise sum of
Lennard-Jones potentials
\begin{equation}
\phi_w(r) = 4\epsilon_w\left[\left(\frac{\sigma_w}{r}\right)^{12} -
  \left(\frac{\sigma_w}{r}\right)^6\right]
\end{equation}
between individual fluid and substrate particles.

$\tilde{w}(r)$ describes the attractive part of the pair potential
between the fluid particles which is taken to be of Lennard-Jones
type, i.e., $\phi_f(r) = 4\epsilon_f[(\sigma_f/r)^{12}-(\sigma_f/r)^6]$.
A division scheme
like the Weeks-Chandler-Andersen (WCA) procedure \cite{wca} provides
an attractive ($\phi_{att}$) and a repulsive 
part ($\phi_{rep}$) as suitable entries into the expression
(\ref{e:functional}). We approximate the attractive part 
by the smooth function
\begin{equation}
\tilde{w}(r) = \frac{4w_0\sigma_f^3}{\pi^2}(r^2+\sigma_f^2)^{-3}
\end{equation}
where 
\begin{equation}\label{e:w0}
w_0 = \int_{{\mathbb R}^3}d^3r\,\tilde{w}(r) = \int_{{\mathbb
R}^3}d^3r\,\phi_{att}(r) = -\frac{32}{9}\sqrt{2}\pi\epsilon_f\sigma_f^3.
\end{equation}
$\tilde{w}(r)$ exhibits the main feature of the attractive van der
Waals interaction, namely the large-distance behavior
$\tilde{w}(r)\sim r^{-6}$.
Its amplitude is chosen such that the integrated strength of
$\tilde{w}(r)$ is the same as the integrated strength of $\phi_{att}$
as constructed from the WCA procedure.
The repulsive part
$\phi_{rep}(r)=\Theta(2^{1/6}\sigma_f-r)(\phi_f(r)+\epsilon_f)$ (with
the Heaviside step function $\Theta$) of the
pair interaction gives rise to a reference 
free energy $f_{HS}(\rho,T)$ of a hard sphere fluid, for which we adopt the
Carnahan-Starling approximation \cite{cs}:
\begin{equation}
f_{HS}(\rho,T) =
k_BT\rho\left(\ln(\rho\lambda^3)-1+\frac{4\eta-3\eta^2}{(1-\eta)^2}\right),
\end{equation}
with the thermal de Broglie wavelength $\lambda$, the dimensionless
packing fraction $\eta=\frac{\pi}{6}\rho(d(T))^3$, and the effective
hard sphere diameter  
\begin{equation}
d(T) = \int\limits_0^{2^{1/6}\sigma_f}dr\,\left\{1-\exp\left(-\frac{\phi_{rep}(r)}{k_BT}\right)\right\}.
\end{equation}
In Eq.~(\ref{e:functional}) the reference free energy is evaluated in
a local density approximation and therefore it does not properly take
into account the details of the packing effects near the wall. If one
would be interested in these aspects, more sophisticated density
functional theories have to be applied. Due to the last term in
Eq.~(\ref{e:functional}) the present density functional is a nonlocal
expression. The bulk phase diagram, i.e., the values for the bulk particle
densities $\rho_l(T,\mu)$ and $\rho_g(T,\mu)$ can be calculated by
minimizing the bulk free energy density
\begin{equation}\label{e:bfedensity}
\Omega_b(\rho,T,\mu) = f_{HS}(\rho,T) + \frac{1}{2}w_0\rho^2 - \mu\rho
\end{equation}
with respect to $\rho$, where $w_0$ is defined in Eq.~(\ref{e:w0}).
Equation~(\ref{e:bfedensity}) follows from
inserting the homogeneous bulk density into
Eq.~(\ref{e:functional}). At two-phase coexistence one has
\begin{equation}
\left.\frac{\partial\Omega_b}{\partial\rho}\right|_{\rho=\rho_g} = 
\left.\frac{\partial\Omega_b}{\partial\rho}\right|_{\rho=\rho_l} = 0
\quad \mbox{and} \quad \Omega_b(\rho_g) = \Omega_b(\rho_l).
\end{equation}

With the bulk properties fixed the functional expression
(\ref{e:functional}) can now be used to 
analyze the morphology and the line tension of a three-phase contact
line. In spite of the relative simplicity of the expressions in
Eq.~(\ref{e:functional}) its full minimization with respect to
$\rho(x,z)$ for the boundary 
conditions described in Fig.~\ref{f:cl_system} represents a big
numerical challenge. Since we are mainly interested in the local
interface position we refrain from seeking this full solution. Instead
we restrict the space of possible density
distributions to a subspace of piecewise constant densities (see
Fig.~\ref{f:cl_sharpkink}). Within this approximation at the
position of the liquid-vapor interface $z=l(x)$ the density varies
steplike between the bulk values determined by
Eq.~(\ref{e:bfedensity}). Moreover, with $\rho(x,z<d_w)=0$ (see
Fig.~\ref{f:cl_sharpkink}) we take into account that there is an 
excluded volume near the substrate surface induced by the repulsive
part of the substrate potential. Approximately one has $d_w =
\frac{1}{2}(\sigma_f+\sigma_w)$. 
Thus the analysis of the grand canonical functional amounts to inserting the
steplike density distribution (``sharp-kink approximation'')
\begin{equation}\label{e:cl_densityansatz}
\hat\rho(x,z) =
\Theta(z-d_w)\{\rho_l\Theta(l(x)-z)+\rho_g\Theta(z-l(x))\}
\end{equation}
into the functional (\ref{e:functional}), with $l(|x|\to\infty)$
asymptotically approaching the function
\begin{equation}
a(x) = l_0\Theta(-x) + (l_0+x\,\tan\theta)\Theta(x);
\end{equation}
$\theta$ is the contact angle (see Fig.~\ref{f:cl_system}).

With this ansatz the grand canonical free energy functional can be
systematically decomposed into bulk, surface, and line
contributions. The decomposition is carried out for a 
finite system, and in a second step the thermodynamic limit is
performed. This lengthy calculation is carried out
in detail in Ref.~\cite{getta}. Here we quote only the main results.
There are artificial surface and line contributions which stem from truncating
the system at finite distances before considering the
thermodynamic limit; we omit them here.
One obtains the following expression for $\Omega[\hat\rho]$:
\begin{eqnarray}\label{e:cl_subdiv}
\Omega[\hat\rho(x,z)] & = & \Lambda^{(l)}(\theta)\,\Omega_b(\rho_l) +
\Lambda^{(g)}(\theta)\,\Omega_b(\rho_g) \nonumber\\
& & + A\,\Omega_s(l_0,\theta) + L_y\,\Omega_l[l(x)],
\end{eqnarray}
with the volumes $\Lambda^{(l)} = \frac{1}{4}L_xL_y\bar{L}(\theta)$
and $\Lambda^{(g)} = \frac{3}{4}L_xL_y\bar{L}(\theta)$, where
$\bar{L}(\theta) = L_z-l_0=\frac{1}{2}L_x\tan(\theta)$, and the surface
area $A = L_xL_y$ (see Fig.~\ref{f:cl_system}). The first two terms in 
Eq.~(\ref{e:cl_subdiv}) describe the
bulk free energies of the liquid and vapor phases, with the bulk free
energy density $\Omega_b(\rho_{\gamma})$ given by Eq.~(\ref{e:bfedensity}). The
surface contribution $\Omega_s(l_0,\theta)$ consists of
the following terms:
\begin{equation}
\Omega_s(l_0,\theta) = l_0\Omega_b(\rho_l) +
\sigma_{wl}+\frac{1}{2}\sigma_{lg} + \frac{1}{2\cos(\theta)}
\sigma_{lg} + \frac{1}{2}\omega(l_0)
\end{equation}
where $\sigma_{wl}$ and $\sigma_{lg}$ are the wall-liquid and
liquid-vapor surface tension, respectively.
This expression corresponds to that for thin liquidlike
wetting films adsorbed on \emph{homogeneous and planar} substrates as
obtained by the same approach \cite{sdreview}. In this context it is shown
that the equilibrium thickness $l_0$ of the liquidlike film minimizes
the effective interface potential 
\begin{equation}\label{e:omega}
\omega(l) = \Delta\rho\left(\rho_l
\int\limits_{l-d_w}^{\infty}dz\,t(z) -
\int\limits_{l}^{\infty}dz\,V(z)\right) =
\sum\limits_{i\geq2}\frac{a_i}{l^i}, \quad l\gg d_w.
\end{equation}
$a_2$ is known as the Hamaker constant. 
The interaction potential $t(z)$ of a fluid particle with a half
space occupied by fluid particles is given by
\begin{equation}
t(z) = \int\limits_z^{\infty} dz'\int_{{\mathbb R}^2}
d^2r_{\parallel} \tilde{w}(\sqrt{r_{\parallel}^2+z'^2}).
\end{equation}
The properties
of the effective interface potential determine the character of the different
wetting transitions. The contact angle $\theta$ follows from Young's
equation:
\begin{equation}
\cos\theta = \frac{\sigma_{wg}-\sigma_{wl}}{\sigma_{lg}} =
1+\frac{\omega(l_0)}{\sigma_{lg}}
\end{equation}
with the liquid-vapor surface tension given within sharp-kink
approximation by
\begin{equation}
\sigma_{lg} = -\frac{1}{2}(\Delta\rho)^2 \int\limits_0^{\infty} dz\,
t(z).
\end{equation}

The line contribution can be split up into one term independent of $l(x)$
and one functionally depending on $l(x)$:
\begin{equation}\label{e:cl_subdivlt}
\Omega_l[l(x)] = \tau_a(l_0,\theta) + \tau_l[l(x)]
\end{equation}
with
\begin{eqnarray}
\tau_a(l_0,\theta) & = &
\frac{w_0}{4\pi}(\Delta\rho)^2\left(1-\frac{\theta}{\tan\theta}\right)
+ \frac{\Delta\rho}{\tan\theta}\left[\left(
-\frac{w_0}{2\pi}\right)\rho_l \right. \\
& &
\left.\times\left\{[(l_0-d_w)^2+\sigma_f]
    \left(-\frac{\pi}{2}+\arctan(l_0-d_w)\right)
+l_0-d_w\right\}+\sum\limits_{i\geq1}\frac{1}{i(i+1)}\frac{u_{i+2}}{l_0^i}\right]
\nonumber
\end{eqnarray}
and
\begin{equation}\label{e:cl_subdivtau}
\tau_l[l(x)] = \tau_{\omega}[l(x)] + \tau_i[l(x)]
\end{equation}
where the first term is given by an integral over the effective
interface potential:
\begin{eqnarray}\label{e:cl_tauomega}
\tau_{\omega}[l(x)] & = & -\Delta\rho\left\{\rho_l
\int\limits_{-\infty}^{\infty} dx \int\limits_{a(x)}^{l(x)} dz \,
(\rho_lt(z-d_w) - V(z))\right\} \nonumber\\
& = & \int\limits_{-\infty}^{\infty} dx \, \{\omega(l(x))-\omega(a(x))\}
\end{eqnarray}
with $\omega(l)$ given by Eq.~(\ref{e:omega}). The expression
\begin{eqnarray}\label{e:cl_nlocfunc}
\tau_i[l(x)] & = & \frac{1}{2}(\Delta\rho)^2 \left\{ 
\int\limits_{-\infty}^{\infty} dx \int\limits_{-\infty}^{\infty} dx'
\int\limits_{a(x)}^{l(x)} dz \int\limits_{-\infty}^{a(x')} dz' \,
\bar{w}(|x-x'|,|z-z'|) \right. \nonumber\\
& & - \left. \int\limits_{-\infty}^{\infty} dx \int\limits_{-\infty}^{\infty} dx'
\int\limits_{a(x)}^{l(x)} dz \int\limits_{l(x')}^{\infty} dz' \,
\bar{w}(|x-x'|,|z-z'|) \right\}
\end{eqnarray}
describes the free energy contribution from the liquid-vapor interface
and is a \emph{nonlocal} functional of $l(x)$. In Eq.~(\ref{e:cl_nlocfunc})
$\bar{w}$ is an integral over the attractive fluid-fluid interaction
\begin{equation}
\bar{w}(x,z) = \int\limits_{-\infty}^{\infty} dy\, \tilde{w}(\sqrt{x^2+y^2+z^2}).
\end{equation}
The expansion of the integrand in Eq.~(\ref{e:cl_nlocfunc}) into a Taylor
series with respect to $|x-x'|$ yields the gradient expansion of the functional
expression, with the leading term
\begin{equation}\label{e:cl_locfunc}
\tau_i^{(loc)}[l(x)] = \sigma_{lg}\int\limits_{-\infty}^{\infty} dx
\left\{ \sqrt{1+\left(\frac{dl}{dx}\right)^2} -
\sqrt{1+\left(\frac{da}{dx}\right)^2} \right\},
\end{equation}
which is now a \emph{local} functional of $l(x)$.

The line contribution (Eq.~(\ref{e:cl_subdivlt})) to the free energy is minimized
by using either Eq.~(\ref{e:cl_nlocfunc}) or Eq.~(\ref{e:cl_locfunc}),
respectively, yielding the equilibrium liquid-vapor interface 
$\bar{l}(x)$ and the corresponding line tension $\tau$ as the minimum
value $\Omega_l[\bar{l}(x)]$. In the following the application of
Eq.~(\ref{e:cl_nlocfunc}) will be called ``nonlocal theory'', whereas
Eq.~(\ref{e:cl_locfunc}) is used for the so-called ``local theory'', also
known as ``interface displacement model''. The expression in 
Eq.~(\ref{e:cl_locfunc}) measures the variation of surface free energy due to
the deformation of the liquid-vapor interface.
Equations (\ref{e:cl_subdivlt})--(\ref{e:cl_tauomega}) and
(\ref{e:cl_locfunc}) provide a 
prescription of how to introduce the parameters serving for a microscopic
description of the underlying molecular interactions into the more
simple interface displacement model which is originally motivated by
phenomenological, macroscopic considerations. 

\subsection{Intrinsic structure of the contact line}

First we analyze the shape of the interface within the local
theory. Minimization of Eq.~(\ref{e:cl_subdivlt}) by using
the functional in Eq.~(\ref{e:cl_locfunc}) leads to solving the
corresponding Euler-Lagrange equation (ELE) following from
$\delta\Omega_l/\delta l(x) = 0$:
\begin{equation}\label{e:cl_eleloc}
\left.\frac{d\omega(l)}{dl}\right|_{l=l(x)} =
-\Delta\rho[\rho_l\,t(l(x)-d_w)-V(l(x))] = 
\frac{\sigma_{lg}l\,''(x)}{(1+(l'(x))^2)^{3/2}} = \sigma_{lg}K(x).
\end{equation}
$K(x)$ is the curvature of
the planar trajectory $(x,l(x))$, with its local radius of curvature
-- which is also one of the two principal radii of curvature of the
surface $(x,y,z=l(x))$ -- given
by $R = 1/K$. It is related to the mean curvature $H$
of the manifold $(x,y,z=l(x))$ by $H(x) = \frac{1}{2}K(x)$.
Equation~(\ref{e:cl_eleloc}) relates the local curvature $K(x)$ to the
derivative of the effective potential (Eq.~(\ref{e:omega})) exerted on the
liquid-vapor interface. The equation is
often referred to as ``augmented Young-Laplace equation'' \cite{kagan}.
This nonlinear ordinary differential equation can easily be
solved using standard numerical tools, e.g., the algorithms from
the Numerical Algorithm Group (NAG). We have used
a Runge-Kutta algorithm with a prescribed starting point $(x_0,l_0)$
and an initial derivative $l_0'$. If $l_0' = 0$, the solution is
$l(x) = l_0 = \mbox{const}$, which describes the thin wetting film of
constant thickness on the homogeneous and planar substrate. If,
however, a very small initial derivative, e.g., $l_0' = 10^{-8}$, is
chosen, then the solution diverges for increasing values
of the lateral coordinate $x$, asymptotically exhibiting the constant
slope $\tan\theta$ for $|x|\to\infty$. The origin of the coordinate system is then
shifted in lateral direction such that the position $x=0$ corresponds
to the intersection of the asymptotes of $l(x)$. 

Figure~\ref{f:cl_local1st}
shows liquid-vapor interfaces around a three-phase contact line
for a system undergoing a first-order wetting
transition. Figure~\ref{f:cl_local1st}(a) displays the temperature
dependence of the profiles $l(x)$, whereas Fig.~\ref{f:cl_local1st}(b)
presents the deviation of the profiles from the asymptotes, i.e., 
$\delta l(x) \equiv l(x)-a(x)$. Apart from a different choice of
interaction parameters and thus a different wetting transition temperature, these 
results are in agreement with those obtained in Ref.~\cite{getta}. One of
the main features in case of a first-order wetting transition is that
$l(x\to\infty)$ approaches its asymptote from below ($\delta l(x>0)<0)$,
whereas $l(x\to -\infty)$ approaches its asymptote from above ($\delta
l(x<0)>0)$ (see also Ref.~\cite{indekeu1994}). A more detailed
discussion of the interface morphology is given in Ref.~\cite{getta}.

Within the nonlocal theory the ELE is the nonlocal integral equation
\begin{equation}\label{e:cl_elenloc}
-\Delta\rho[\rho_l\,t(l(x)-d_w)-V(l(x))] =
-(\Delta\rho)^2\int\limits_{-\infty}^{\infty}dx'
\int\limits_0^{l(x')-l(x)} dz' \, \bar{w}(|x-x'|,|z'|),
\end{equation}
i.e., the right-hand side of the local differential ELE (\ref{e:cl_eleloc})
is replaced by an integral expression.
In Ref.~\cite{getta} the interface profiles within the nonlocal
theory have been obtained from a numerical solution of the ELE
(\ref{e:cl_elenloc}). The insertion of the discretization
$(x_i,l_i=l(x_i))$ of the function $l(x)$ on a lattice with $N$ mesh
points into Eq.~(\ref{e:cl_elenloc}) yields a set of
$N$ coupled equations for the values $l_i$ which can be
solved using standard numerical tools. This in itself is already a demanding
task compared to the relatively easy numerical solution of the local
ELE (\ref{e:cl_eleloc}). But moreover the solution distinctly
depends on the number of mesh points $N$, a fact that due to low
computer power has not been revealed in Ref.~\cite{getta}. An
enhancement of $N$ such that the numerical resolution is satisfactory
would require large computer memory and a very long time of computation.
For this reason we numerically minimize the functional
$\Omega_l^{(nloc)}[l(x)]$ itself, i.e., Eq.~(\ref{e:cl_subdivlt}) with the
nonlocal expression (\ref{e:cl_nlocfunc}) for $\tau_i[l(x)]$, instead
of solving the ELE. To this end the interface is also
discretized $(x_i,l_i = l(x_i))$ with a linear interpolation between
the mesh points. The $z$ integration in
Eq.~(\ref{e:cl_tauomega}) and the $z$ and $z'$ integrations in
Eq.~(\ref{e:cl_nlocfunc}) can be carried out analytically. With the remaining
integrations performed numerically,
the discretization procedure yields a function $\Omega_l(\{l_i\})$
which is minimized with respect to the $N$ variables $l_i$
using a particulary suitable algorithm from the NAG routine library.
The price of a significant increase of the numerical effort, which is
required by this approach as opposed to the seemingly easier way of
solving Eq.~(\ref{e:cl_elenloc}) directly, is
justified by the fact that the solution does
not depend sensitively on the choice of the
mesh size $\delta x = x_{i+1}-x_i$, as it does for the ELE
(\ref{e:cl_elenloc}). A repetition of the numerical minimization using
a mesh with a larger number of points $N$ yields an identical
result, which corroborates the reliability of the minimization
procedure. Moreover, our results are in accordance with those obtained
for geometrically structured, wegde-shaped systems which also indicate that
interface profiles obtained from the local and the nonlocal theory differ
only slightly \cite{boigs}.

The same minimization procedure can be applied to check the accuracy of
the calculation of interface profiles within the local
theory. As expected, the minimization of the expression
(\ref{e:cl_subdivlt}) using the local functional
Eq.~(\ref{e:cl_locfunc}) yields results identical to those 
obtained from the solution of the differential equation
(\ref{e:cl_eleloc}) whose numerical solutions do not depend on the number
of mesh points.

For the same system as discussed in Fig.~\ref{f:cl_local1st},
Fig.~\ref{f:cl_localnlocal1st} displays typical examples for the
difference between the predictions obtained from the local and the 
nonlocal theory in terms of the deviation from the asymptotes $\delta
l(x)$. Figure~\ref{f:cl_localnlocal_curv} shows the curvature $K(x)$
of the respective interfaces. $K(x)$ is
slightly underestimated within the local theory.
These figures clearly demonstrate that the obvious differences between the
predictions of both theories are small and confined to those regions where $K(x)$
is largest. Significant deviations of the nonlocal predictions from the
local ones only occur at sections of the interface where the radius of
curvature $1/K(x)$ is less than $10\sigma_f$.
This corrects the nonlocal results obtained in Ref.~\cite{getta} from solving the
ELE. Our refined
numerical analysis shows that the differences are largest for high
values of contact angle, i.e., for temperatures far below the wetting
temperature. Upon approaching
the wetting transition the contact angle and the curvature of the
interface decrease, and the difference between the local and the nonlocal
predictions becomes smaller.

Figure~\ref{f:cl_localcrit} displays liquid-vapor interfaces
calculated within the local theory for a system which exhibits
a critical wetting transition at $T_w^* = 1.2$. In the case of
critical wetting $l(x)$ approaches the asympotes from above both for
$x\to\infty$ and for $x\to-\infty$, i.e., $\delta l(x)$ is always
positive. If the system undergoes a critical wetting transition,
the contact angles are rather small, and the interface profile exhibits
very small local curvatures (see Fig.~\ref{f:cl_localcrit}) and an
extremely broad transition region
(which is the interval where the deviation of $l(x)$ from its asymptotes
is larger than a certain prescribed value).

In the case of a critical wetting transition it is presently impossible to
perform a full numerical minimization of $\Omega_l^{(nloc)}$: due to
the huge required system sizes over which the numerical integrations
in Eqs.~(\ref{e:cl_tauomega}) and (\ref{e:cl_nlocfunc}) have to be
performed the available computer resources are by far exceeded. However,
for a given equilibrium profile $\bar{l}(x)$ calculated
within the local theory the relative difference
$(\hat{\Omega}_l^{(nloc)}-\Omega_l^{(loc)})/\Omega_l^{(loc)}$, with
$\hat{\Omega}_l^{(nloc)}$ evaluated for $\bar{l}(x)$, is of the order of
$10^{-3}$. Therefore in the case of critical wetting we expect
the difference between the interface profiles (as well as the line
tensions) calculated within the local and the nonlocal theory to be negligibly
small.

\subsection{Line tension}

The line tension $\tau$ corresponding to the equilibrium liquid-vapor
interface profile follows from Eqs.~(\ref{e:cl_subdivlt})-(\ref{e:cl_locfunc}):
\begin{equation}
\tau^{((n)loc)} = \min_{\{l(x)\}}\Omega_l^{((n)loc)}[l(x)] =
\Omega_l^{((n)loc)}[\bar{l}(x)].
\end{equation}
Within the local theory $\tau^{(loc)}$ is obtained
by inserting the solution $\bar{l}(x)$ of the ELE into
$\Omega_l^{(loc)}$; within the nonlocal theory $\tau^{(nloc)}$
immediately follows from the minimization procedure itself.

Figure~\ref{f:cl_lt} shows the results for $\tau^{(loc)}$ and
$\tau^{(nloc)}$ for the system undergoing a first-order wetting
transition at $T_w^* = 1.102$ (compare
Figs.~\ref{f:cl_local1st}-\ref{f:cl_localnlocal_curv}). Due to the
large numerical effort only a few results have
been obtained in the nonlocal theory. The data indicate that
the difference $\tau^{(nloc)}-\tau^{(loc)}$ is rather small
and vanishes upon approaching the wetting transition.
A thorough discussion of the different
contributions to the line tension is presented in
Ref.~\cite{getta}, including a comparison with the singular behavior
predicted by Indekeu \cite{indekeu1992,indekeu1994}. The results
obtained in Ref.~\cite{getta} for the local theory are still valid; therefore we
can refrain from a further discussion here. We only show in
Fig.~\ref{f:cl_contriblt} the temperature behavior of the two contributions
$\tau_{\omega}$ and $\tau_i$
(Eqs.~(\ref{e:cl_subdivtau})-(\ref{e:cl_locfunc})) to the line tension
$\tau$, as derived within local and nonlocal theory. The difference
$\tau_i^{(nloc)}-\tau_i^{(loc)}$ is about twice the difference
$\tau_{\omega}^{(nloc)}-\tau_{\omega}^{(loc)}$, and both vanish for
$T \to T_w$.

\section{Adsorption on a chemical step}
\label{s:chemicalstep}

\subsection{Model and density functional theory}

The study of the system shown in Fig.~\ref{f:step_system} is a first
step towards the understanding of wetting on
chemically heterogeneous surfaces \cite{koch}. In this case the
translational symmetry in lateral directions is broken by the
variation of the substrate potential in $x$ direction
(Fig.~\ref{f:step_system}). We describe the fluid in the half space
$\{{\mathbf r}\in {\mathbb R}^3|z>0\}$ as in Sec.~\ref{s:contactline}
(Eq.~(\ref{e:functional})). The only difference is that the substrate
potential $V(z)$ in Eq.~(\ref{e:substpothom}) is replaced by
$V(x,z)$. As boundary condition one has $\rho(x,z\to\infty) = \rho_g$
for \emph{all} $x$, including $|x|\to\infty$. To be specific we assume
that the substrate potential is the pairwise superposition of
Lennard-Jones pair potentials between the fluid and substrate particles,
\begin{equation}
\phi_{\pm}(r) =
4\epsilon_{\pm}\left[\left(\frac{\sigma_{\pm}}{r}\right)^{12} -
  \left(\frac{\sigma_{\pm}}{r}\right)^6\right],
\end{equation}
where the $+$ and $-$ signs stand for interaction of a fluid particle
with the different species occupying the quarter spaces $w_+ = \{{\mathbf r}\in
{\mathbb R}^3|x>0,z<0\}$ and $w_- = \{{\mathbf r}\in {\mathbb R}^3|x<0,z<0\}$,
respectively. This superposition yields \cite{koch}
\begin{eqnarray}
V_{att}(x,z) & = & -\frac{u_3^++u_3^-}{2}\frac{1}{z^3} +
\frac{u_3^+-u_3^-}{2}
\left(\frac{1}{x^3}-\left(\frac{r}{xz}\right)^3+\frac{3}{2}\frac{1}{xzr}\right)
\nonumber\\
& & -\frac{u_{4,z}^++u_{4,z}^-}{2}\frac{1}{z^4} -
\frac{u_{4,z}^+-u_{4,z}^-}{2}
\left(\frac{x}{z^4r}+\frac{1}{2}\frac{x}{z^2r^3}\right) \nonumber\\
& & + \frac{u_{4,x}^+-u_{4,x}^-}{2}
\left(\frac{1}{x^4}-\left(\frac{z}{x^4r}+\frac{1}{2}\frac{z}{x^2r^3}\right)\right)
+ {\mathcal O}(x^{-m}z^{-n},\quad m+n\geq5)
\end{eqnarray}
for the attractive part of the substrate potential, where $r\equiv
r(x,z) = \sqrt{x^2+z^2}$. The coefficients $u_3^{\pm}$,
$u_{4,z}^{\pm}$, and $u_{4,x}^{\pm}$ are functions of the interaction
potential parameters $\epsilon_{\pm}$ and $\sigma_{\pm}$; $u_3^{\pm}$
and $u_{4,z}^{\pm}$ are the coefficients of the corresponding
homogeneous, flat, semi-infinite substrate $w_{\pm}$ occupied by
the species ``$\pm$'' with the
substrate potential of a homogeneous substrate defined as in
Eq.~(\ref{e:substpothom}). For the repulsive contribution to the
substrate potential we use the simple ansatz
\begin{equation}
V_{rep}(x,z) = \Theta(-x)\frac{u_9^-}{z^9} +
\Theta(x)\frac{u_9^+}{z^9},
\end{equation}
because the repulsive interaction of each half of the substrate
decays rapidly.

This system has been studied in detail in Ref.~\cite{koch}. As stated
in the Introduction we revisit this problem in order to study the
interface morphology with the refined numerical analysis described in
Sec.~\ref{s:contactline}. Moreover, in Ref.~\cite{koch} only the
cases of critical and complete wetting have been covered. Here we also
study a first-order wetting transition for which one can expect the
largest differences between the local and the nonlocal approach
(see Sec.~\ref{s:contactline}).

The behavior of the substrate potential $V(x,z)$ for large $|x|$ and
$z$ fixed, i.e., far from the heterogeneity,
\begin{equation}
V(x\to\pm\infty,z) = -\frac{u_3^{\pm}}{z^3} -
\frac{u_{4,z}^{\pm}}{z^4} + \frac{u_9^{\pm}}{z^9} 
\pm \frac{u_3^+-u_3^-}{|x|^3} + {\mathcal O}(|x|^{-4}),
\end{equation}
causes the liquid-vapor interface to asymptotically approach the
constant values $l(x\to\pm\infty) = l_{\pm}$ valid for the
corresponding homogeneous substrate.
Therefore it is appropriate to make the following sharp-kink
ansatz for the particle density distribution
(see Eq.~(\ref{e:cl_densityansatz})):
\begin{eqnarray}
\hat\rho(x,z) & = &
[\Theta(-x)\Theta(z-d_w^-)+\Theta(x)\Theta(z-d_w^+)] \nonumber\\
& & \times[\rho_l\Theta(l(x)-z)+\rho_g\Theta(z-l(x))].
\end{eqnarray}
For this density distribution the grand canonical free energy
functional decomposes into bulk, surface and
line contributions such that the surface contribution
describes the wetting of each single homogeneous, flat,
semi-infinite substrate:
\begin{eqnarray}
\Omega([\hat\rho(x,z)];T,\mu;[\tilde{w}],[V]) & = &
\Lambda\,\Omega_b(\rho_g,T,\mu)
+ A\,\Omega_s(l_{\pm};T,\mu;[\tilde{w}],[V])
\nonumber\\
& & + L_y\,\Omega_l([l(x)];T,\mu;[\tilde{w}],[V]),
\end{eqnarray}
where $\Omega_b$ is given by Eq.~(\ref{e:bfedensity}), $\Lambda =
L_xL_yL_z$ is the volume filled with fluid, and $A = L_xL_y$ is the
surface area of the substrate surface (see Fig.~\ref{f:step_system}).
Again artificial terms due to truncations are omitted.
\begin{equation}
\Omega_s(l_{\pm}) = \frac{1}{2}(l_++l_-)\Delta\Omega_b +
\sigma_{lg} + \frac{1}{2}(\sigma_{wl}(l_+)+\sigma_{wl}(l_-)) +
\frac{1}{2}(\omega(l_+)+\omega(l_-))
\end{equation}
is the arithmetic mean of the surface free energy densities
corresponding to wetting of the homogeneous substrates $w_+$ and
$w_-$ by liquidlike layers of thickness $l_+$ and $l_-$,
respectively. $\Delta\Omega_b = \Delta\rho\,\Delta\mu+{\mathcal
  O}((\Delta\mu)^2)$ with $\Delta\rho=\rho_l-\rho_g$ measures the
undersaturation if the liquid phase is not yet thermodynamically
stable, i.e., if $\Delta\mu = \mu_0-\mu >0$.
The line contribution $\Omega_l[l(x)]$ reads
\begin{equation}\label{e:step_subdivlt}
\Omega_l[l(x)] = \tau_{wl} + \tau_{lg} + \tilde{\omega}[l(x)]
\end{equation}
with
\begin{eqnarray}
\tilde{\omega}[l(x)] & = &
\Delta\Omega_b\int\limits_{-\infty}^{\infty} dx (l(x)-l_{\infty}(x)) +
\Delta\rho\,\rho_l\int\limits_{-\infty}^{\infty} dx
\int\limits_{l(x)-d_w^+}^{l_{\infty}(x)-d_w^+}dz \, t(z)
-\Delta\rho\int\limits_{-\infty}^{\infty} dx
\int\limits_{l(x)}^{l_{\infty}(x)}dz \, V(x,z)\nonumber\\
& & - \Delta\rho\,\rho_l \left( \int\limits_{-\infty}^{\infty} dx
\int\limits_{l(x)-d_w^+}^{l_{\infty}(x)-d_w^+}dz \, \bar{t}(x,z) - 
\int\limits_{-\infty}^{\infty} dx
\int\limits_{l(x)-d_w^-}^{l_{\infty}(x)-d_w^-}dz \, \bar{t}(x,z) \right)
\nonumber\\
& & - \frac{1}{2} (\Delta\rho)^2 \int\limits_{-\infty}^{\infty} dx
\int\limits_{-\infty}^{\infty} dx' \int\limits_0^{\infty} dz
\int\limits_0^{l(x)-l(x')} dz' \, \bar{w}(|x-x'|,|z-z'|)
\end{eqnarray}
where $l_{\infty}(x) \equiv \Theta(-x)l_-+\Theta(x)l_+$ and
\begin{displaymath}
\bar{t}(x,z) \equiv \int\limits_x^{\infty}dx'\int\limits_z^{\infty}dz'
\, \bar{w}(x',z').
\end{displaymath}
The rather lengthy expressions $\tau_{wl}$ and $\tau_{lg}$ do not
depend on the function $l(x)$ and are given in Ref.~\cite{koch}.
$\tilde{\omega}[l(x)]$ is a nonlocal functional which has
to be minimized to yield the equilibrium profile $\bar{l}(x)$ and the line
tension $\tau = \Omega_l[\bar{l}(x)]$.

The local approximation of the quadruple integral in $\tilde{\omega}[l(x)]$ --
which describes the free energy contribution due to deformation of the
liquid-vapor interface -- is given by the local functional
\begin{eqnarray}
- \frac{1}{2} (\Delta\rho)^2 & & \int\limits_{-\infty}^{\infty} dx
\int\limits_{-\infty}^{\infty} dx' \int\limits_0^{\infty} dz
\int\limits_0^{l(x)-l(x')} dz' \, \bar{w}(|x-x'|,|z-z'|) \nonumber\\
& & \longrightarrow \quad \sigma_{lg}\int\limits_{-\infty}^{\infty}dx
\left\{\sqrt{1+\left(\frac{dl}{dx}\right)^2} - 1 \right\}.
\end{eqnarray}

\subsection{Numerical results for the interface morphology}

Within the local theory the interface morphology follows from the ELE
\begin{equation}
\Delta\Omega_b-\Delta\rho[\rho_l\,t(l(x)-d_w)-V(x,l(x))] =
\frac{\sigma_{lg}l\,''(x)}{(1+(l'(x))^2)^{3/2}} = \sigma_{lg}K(x).
\end{equation}
Apart from the finite undersaturation $\Delta\Omega_b\geq 0$ and the
different substrate potential the structure of this ELE is the same as
in Eq.~(\ref{e:cl_eleloc}). For reasons of simplicity we
assume that $d_w \equiv d_w^+ = d_w^-$. Guided by the experience from
Sec.~\ref{s:contactline} the equilibrium profile within the nonlocal
theory is obtained by minimizing $\Omega_l$
(Eq.~(\ref{e:step_subdivlt})) instead of solving the corresponding ELE
as it has been done in Ref.~\cite{koch}. Here we apply the same
numerical procedures as in Sec.~\ref{s:contactline}.

Figure~\ref{f:step_complete} displays interface profiles on a complete
wetting path, i.e., along an isotherm $(T=\mbox{const},
\Delta\mu\to0)$; the system exhibits critical wetting
transitions at $T_w^* = k_BT_w/\epsilon_f = 1.2$ on the substrate
$w_-$ and at $T_w^* = 1.0$ on the substrate $w_+$, respectively. For
the choice $T^*=1.1$ at coexistence the substrate $w_+$ is completely
wet whereas $w_-$ is only partially wet. We find that the difference
between the local and the nonlocal 
results for $l(x)$ as well as for the line tension $\tau$
is smaller than the numerical resolution and therefore not
significant. The minimization procedure requires a large numerical
effort and is only applied for values of $\mu$ not too close to
coexistence $\mu = \mu_0$ such that the effective size of the system,
which has to be treated numerically, is not too large.

For larger effective system sizes, i.e., close to coexistence $\Delta\mu=0$, we
have to resort to an alternative procedure for evaluating the
equilibrium interface profile $\bar{l}(x)$ within the nonlocal
theory. The corresponding ELE
\begin{equation}
\Delta\Omega_b
-\Delta\rho[\rho_l\,t(l(x)-d_w)-V(x,l(x))] =
-(\Delta\rho)^2\int\limits_{-\infty}^{\infty}dx'
\int\limits_0^{l(x')-l(x)} dz' \, \bar{w}(|x-x'|,|z'|)
\end{equation}
(compare Eq.~(\ref{e:cl_elenloc})) is solved numerically for a
certain choice of thermodynamical parameters $(T,\mu)$. To this end
the lateral extension of the system is truncated and $l(x)$ is
discretized on a lattice with $N$ lattice points and a mesh size $\delta x$, 
yielding a set of $N$ coupled equations for the values $l_i = l(x_i)$.
These equations are solved using a suitable numerical algorithm. The
solutions depend on the mesh size $\delta x$ in such a way
that solutions $l(x;\delta x)$ for
different $\delta x$ differ only with respect to a simple rescaling of
the $x$ axis by a constant numerical factor. We note that this
particular scaling behavior of the numerical solutions of the ELE only
occurs in the present case of a substrate with a chemical step but not
for the intrinsic structure of the contact lines investigated in
Sec.~\ref{s:contactline}. Thus in the latter case this alternative
approach is not applicable. The procedure of finding the correct solution is as
follows. We calculate solutions of the ELE
on lattices with different $\delta x$. A suitable
length scale is defined which measures the typical width $\Delta$ on which the
profile varies between its asymptotic values $l_+$ and $l_-$, e.g.,
the distance between the points where $l(x)$ deviates by $10\%$ from
the asymptotes. As a function of $\delta x$ the width $\Delta$ asymptotically 
approaches a linear dependence $\Delta(\delta x\to 0) = A\,\delta x + B$
which can be extrapolated to $\delta x = 0$. This yields
a factor $\gamma_0 = \Delta(\delta x = 0)/\Delta(\delta x_0)$ by which
a certain solution $l(x;\delta x_0)$ of the ELE has to be
scaled in $x$ direction in order to find the limiting profile
corresponding to an infinitely fine lattice
(see Fig.~\ref{f:step_ndependence}). In order to test the reliability
of this approach we have carried out the following additional
cross-check: a solution $l(x;\delta x_0)$ of the ELE
obtained by using a mesh size $\delta x_0$ is scaled in
$x$ direction by different factors, i.e., $x \mapsto \gamma \, x$;
the correct scaling factor $\gamma_0$ is the one for which the line
contribution to the free energy $\Omega_l[l(\gamma \, x;\delta
x_0)]$ is minimized. Both 
procedures are based on the assumption that the true solution differs from
each of the numerically obtained solutions $l(x;\delta x)$ only by a simple
rescaling of the $x$ axis. (For those systems, for which the full
minimization and the solution of the ELE can be carried out both, this
assumption has been verified.)
Within the numerical accuracy the two methods give identical
results for the scaling factor. Moreover, as expected the solution
obtained by rescaling of the $x$ axis with this optimal factor is
indistinguishable from the one predicted by the local theory.

Figure~\ref{f:step_critical} shows interface profiles for a thermodynamic
path along coexistence $(T \to T_w, \Delta\mu=0)$. The system
exhibits critical wetting at the same transition temperature $T_w^* =
1.0$ for both substrates $w_+$ and $w_-$; the parameters of the
substrate potential are chosen such that the
asymptotic equilibrium film thicknesses $l_+$ and $l_-$ are different
although the wetting transition temperatures are equal.
Due to the broad crossover region these profiles have been obtained by
using the scaling procedure as described for finding the minimum of $\Omega_l$.
Also in this case the local and the nonlocal results are
indistinguishable. Not only the profiles but also the line tensions
obtained within the local and the nonlocal theory, respectively, are
very close. The relative difference of the line tensions
$(\tau^{(nloc)}-\tau^{(loc)})/\tau^{(loc)}$ is of the order of
$10^{-4}$, both for complete as well as for critical wetting.

In Fig.~\ref{f:step_1storder} we present results for
interface profiles for a system with very thin liquidlike wetting films.
Both corresponding homogeneous substrates undergo first-order wetting
transitions: $w_+$ at the temperature $T_w^* \approx 1.102$ and $w_-$
at $T_w^* \approx 1.314$; both wetting temperatures are above the temperatures
considered here. In the previous examples
the differences between the results obtained within the local and
nonlocal theory are small. But here they are detectable since the
interface curvatures are larger than in a system 
with thick wetting films as they occur for critical and complete
wetting. The relative difference 
$(\tau^{(nloc)}-\tau^{(loc)})/\tau^{(loc)}$ is of the order of $10^{-2}$.

\section{Summary}
\label{s:summary}

\begin{figure}
\begin{center}
\epsfig{file=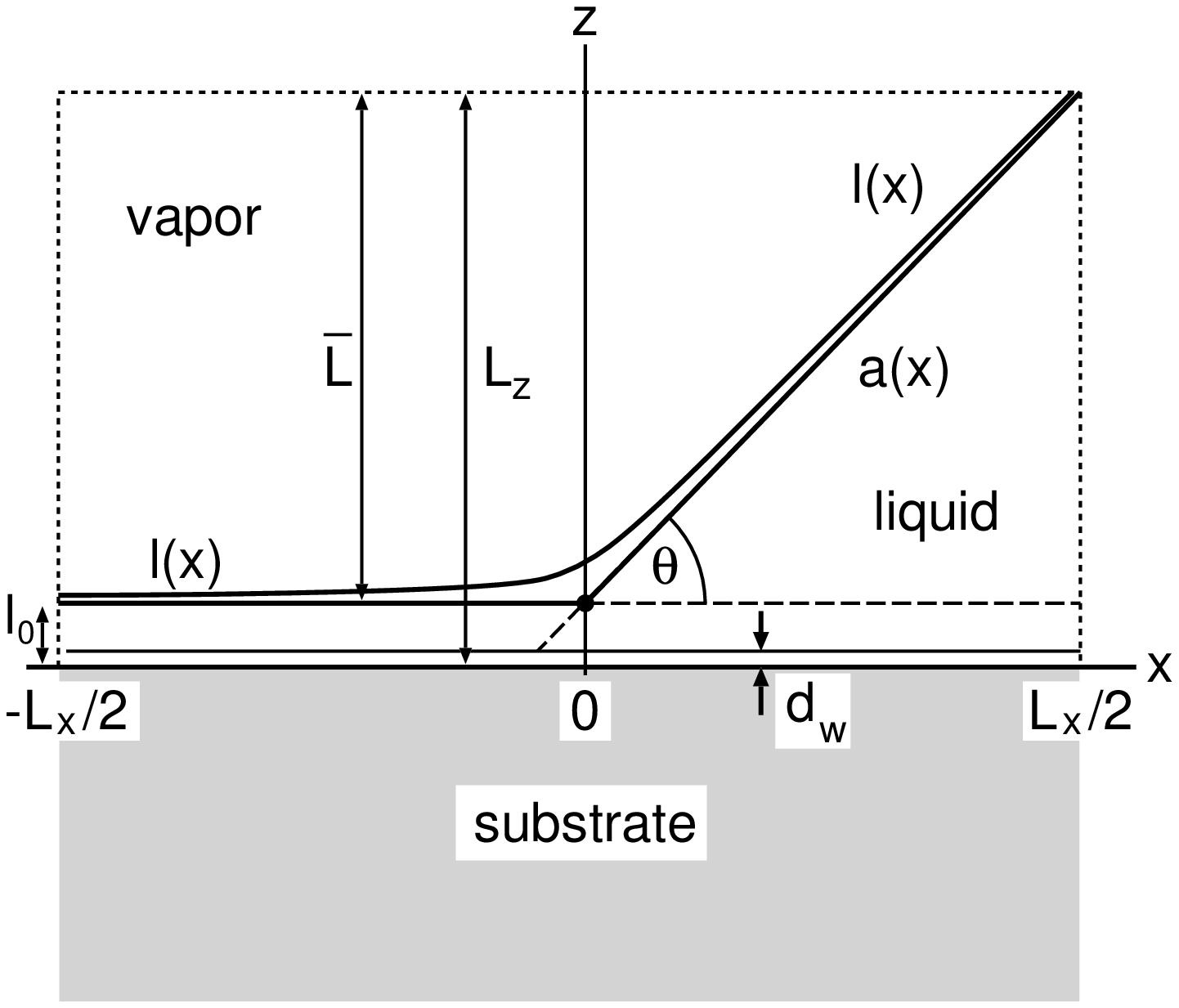, width=13cm}
\end{center}
\caption{\label{f:cl_system}
Shape $l(x)$ of the liquid-vapor interface near the
three-phase contact line on a homogeneous, planar substrate whose
surface is located at $z=0$. $l_0$
is the thickness of the microscopic liquid wetting film, $d_w$ denotes
an excluded volume due to the repulsive part of the substrate
potential. $a(x)$ is the asymptote $l(x)$ 
in the limit $|x|\to\infty$. The angle $\theta$ between $a(x>0)$ and
the substrate surface is the contact angle given by Young's law
$\cos\theta = (\sigma_{wg}-\sigma_{wl})/\sigma_{lg}$. The intersection
between the asymptotes $a(x<0)$ and $a(x>0)$ defines the position
$x=0$. In order to facilitate the proper thermodynamic limit the
system is truncated at $x=\pm L_x/2$ and $z=L_z$. The
configuration is taken to be translationally invariant in the $y$
direction.}
\end{figure}

\begin{figure}
\begin{center}
\epsfig{file=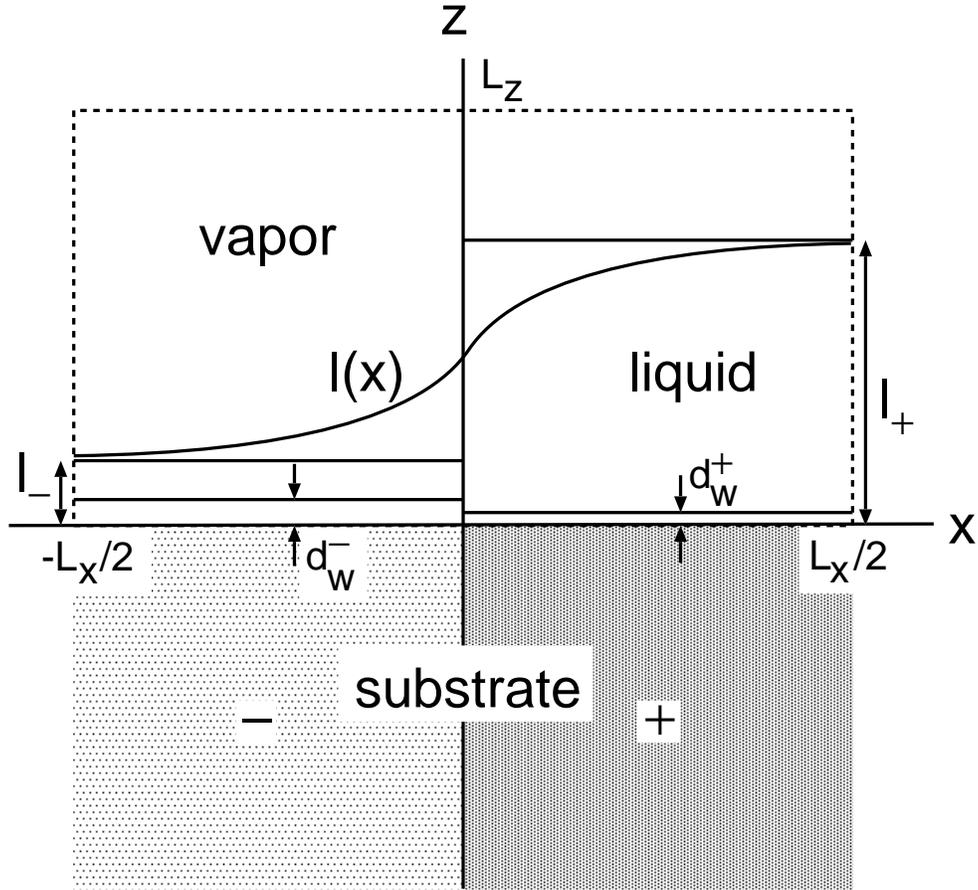, width=13cm}
\end{center}
\caption{\label{f:step_system}
Morphology $l(x)$ of a wetting film which covers a planar substrate
with surface $z=0$. The substrate consists of two halves meeting at
$x=0$. The material filling the left (right) half favors a thin
(thick) wetting film. $l_{\mp} = l(x\to\mp\infty)$ are the equilibrium
film thicknesses of the corresponding homogeneous substrates, which
are characterized also by different excluded volumes $d_w^{\mp}$. The
system is translationally invariant in the $y$ direction. It is
truncated at $z=L_z$ and $x=\mp L_x/2$ in order to facilitate the
proper thermodynamic limit.}
\end{figure}

\begin{figure}
\begin{center}
\epsfig{file=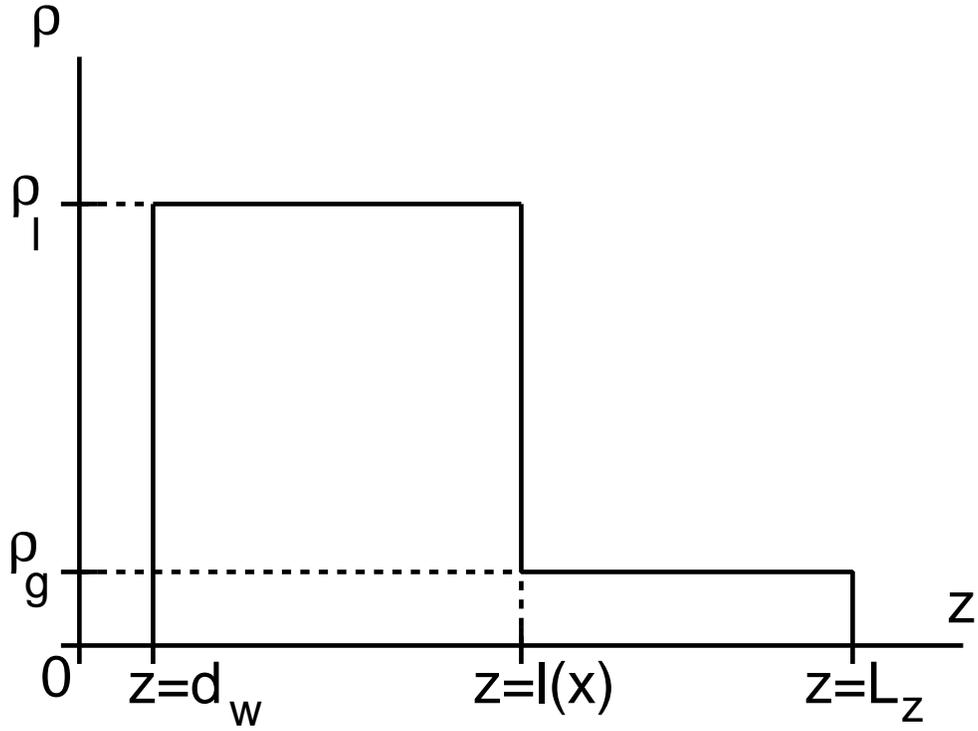, width=13cm}
\end{center}
\caption{\label{f:cl_sharpkink}
Sharp-kink approximation for the particle density distribution. At the
position $z=l(x)$ of the liquid-vapor interface the particle density
distribution varies steplike between the
constant liquid ($\rho_l$) and the constant gas density ($\rho_g$). At
$z=L_z$ the density is truncated in order to facilitate the
thermodynamic limit. Moreover, the density vanishes for $z<d_w$ 
due to the repulsion between the fluid and the substrate particles.}
\end{figure}

\begin{figure}
\begin{center}
\epsfig{file=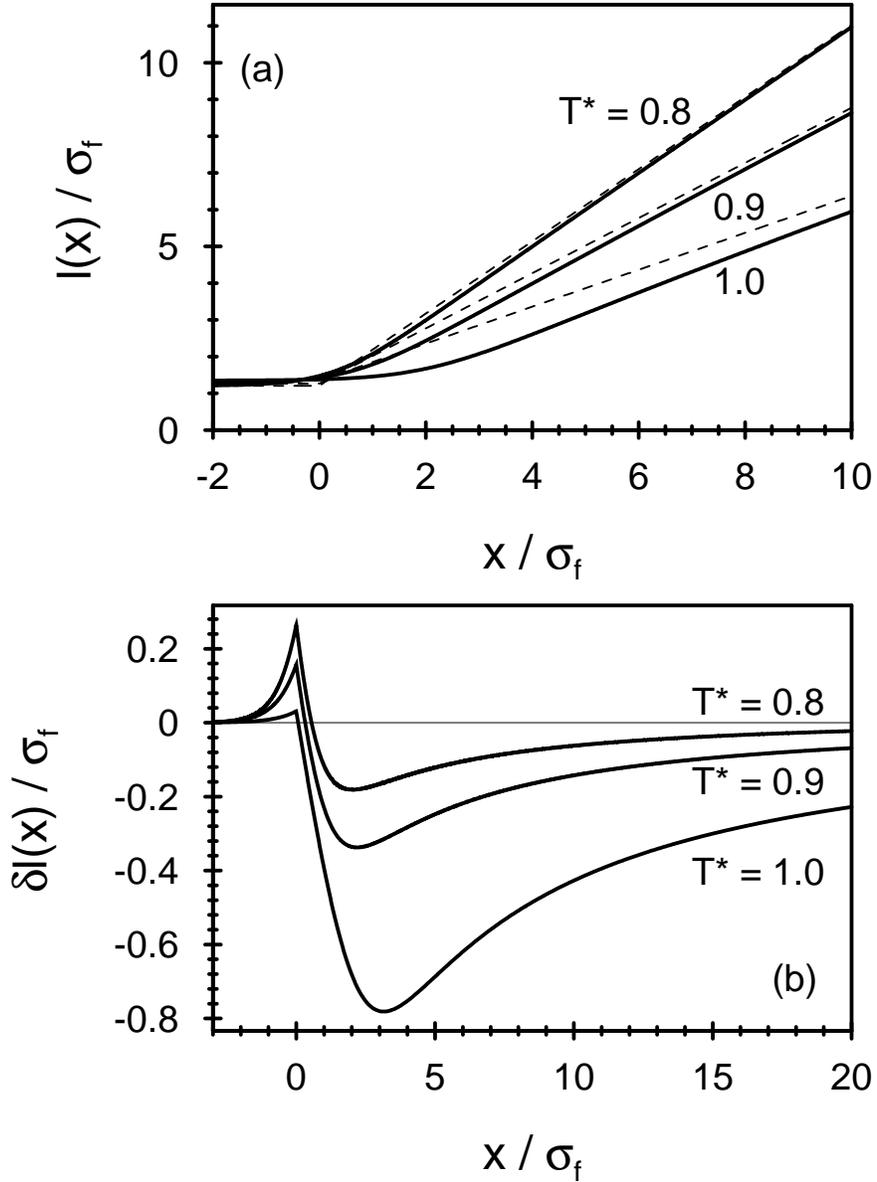, width=11cm, bbllx=80, bblly=190,
  bburx=520, bbury=820}
\end{center}
\caption{\label{f:cl_local1st}
Temperature dependence of the shape of the liquid-vapor interface
calculated within the local theory for a system
which exhibits a first-order wetting transition at $T_w^* =
k_BT_w/\epsilon_f \approx 1.102$. The substrate potential parameters are
$d_w=1.05\sigma_f$, $u_3=3.710\epsilon_f\sigma_f^3$,
$u_4=5.566\epsilon_f\sigma_f^4$, and 
$u_9=0.876\epsilon_f\sigma_f^9$. In (a) the interface profiles $l(x)$
(full lines) and their asymptotes $a(x)$ (dashed lines)
are shown, whereas (b) displays the deviation $\delta l(x) =
l(x)-a(x)$ of $l(x)$ from its asymptotes. The corresponding contact
angles are $\theta \approx 44.5^{\circ}$, $\theta \approx 36.9^{\circ}$,
and $\theta \approx 26.6^{\circ}$ for $T^* = 0.8$, $0.9$, and $1.0$,
respectively. $l(x)$ approaches its asymptotes from above for
$x\to-\infty$ and from below for $x\to\infty$. $l(x)$ closely follows
the asymptotes for large values of the contact angle, i.e., for low
temperatures.}
\end{figure}

\begin{figure}
\begin{center}
\epsfig{file=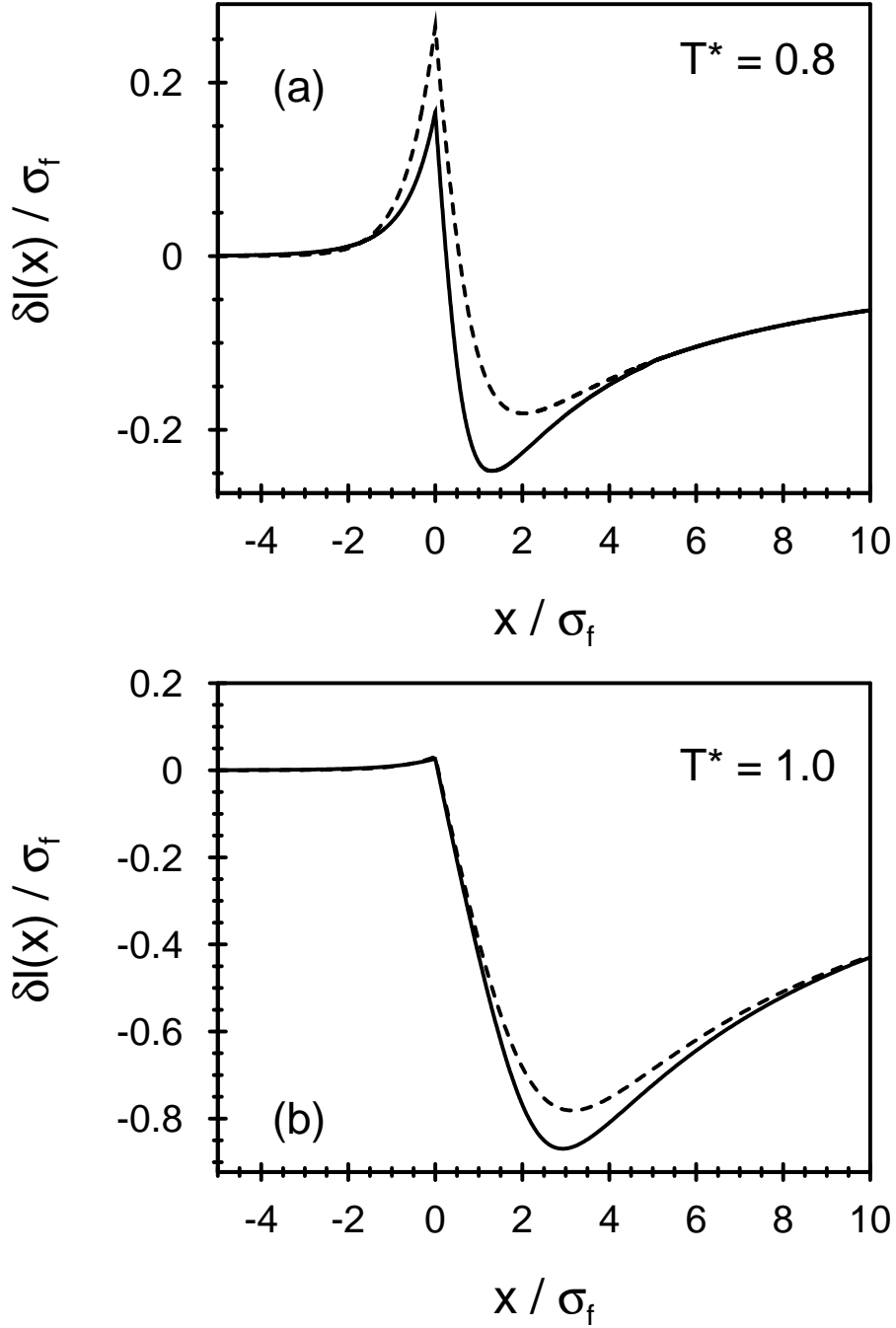, width=11cm, bbllx=95,
  bblly=155, bburx=520, bbury=840}
\end{center}
\caption{\label{f:cl_localnlocal1st}
Comparison between liquid-vapor interface profiles calculated within the
local (dashed lines) and the nonlocal (full lines) theory for the
same system as discussed in
Fig.~\ref{f:cl_local1st}, in terms of the difference 
$\delta l(x) = l(x)-a(x)$ for $T^*=0.8$ (a) and $T^*=1.0$ (b). The
difference between the local and 
the nonlocal results decreases upon approaching the wetting
transition. The local theory slightly underestimates the local curvature $K(x)$.}
\end{figure}

\begin{figure}
\begin{center}
\epsfig{file=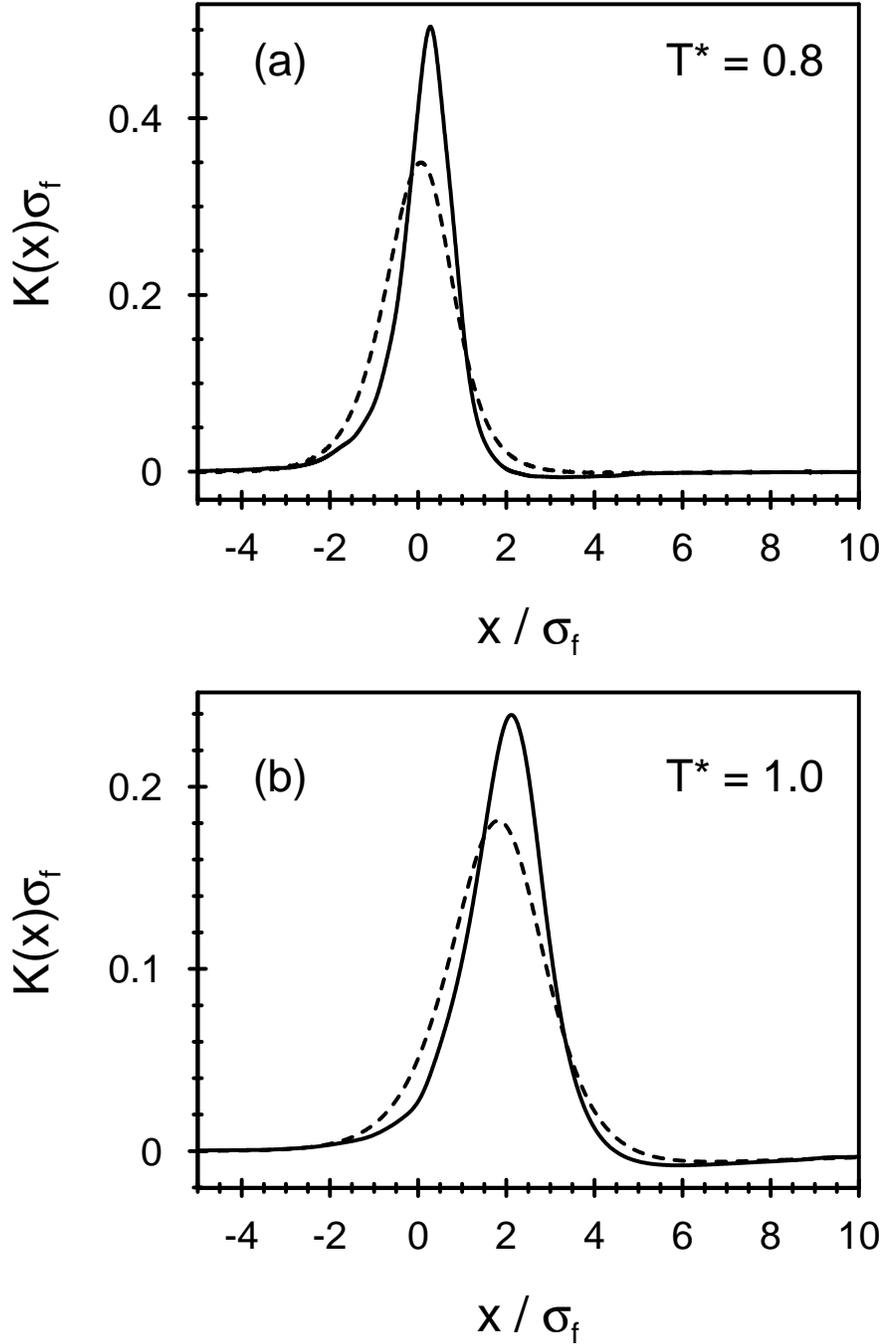, width=11cm, bbllx=95,
  bblly=155, bburx=520, bbury=840}
\end{center}
\caption{\label{f:cl_localnlocal_curv}
Comparison between the curvatures $K(x) =
l\,''(x)/(1+(l\,'(x))^2)^{3/2}$ of the liquid-vapor interface profiles
as calculated within the local (dashed lines) and the nonlocal (full
lines) theory for the same system as in Fig.~\ref{f:cl_local1st}.
The parameters chosen for (a) and (b) are the same as for
Figs.~\ref{f:cl_localnlocal1st}(a) and \ref{f:cl_localnlocal1st}(b),
respectively. Also for the curvature the difference between 
the local and the nonlocal results decreases upon approaching the wetting
transition. The local theory slightly underestimates the curvature.}
\end{figure}

\begin{figure}
\begin{center}
\epsfig{file=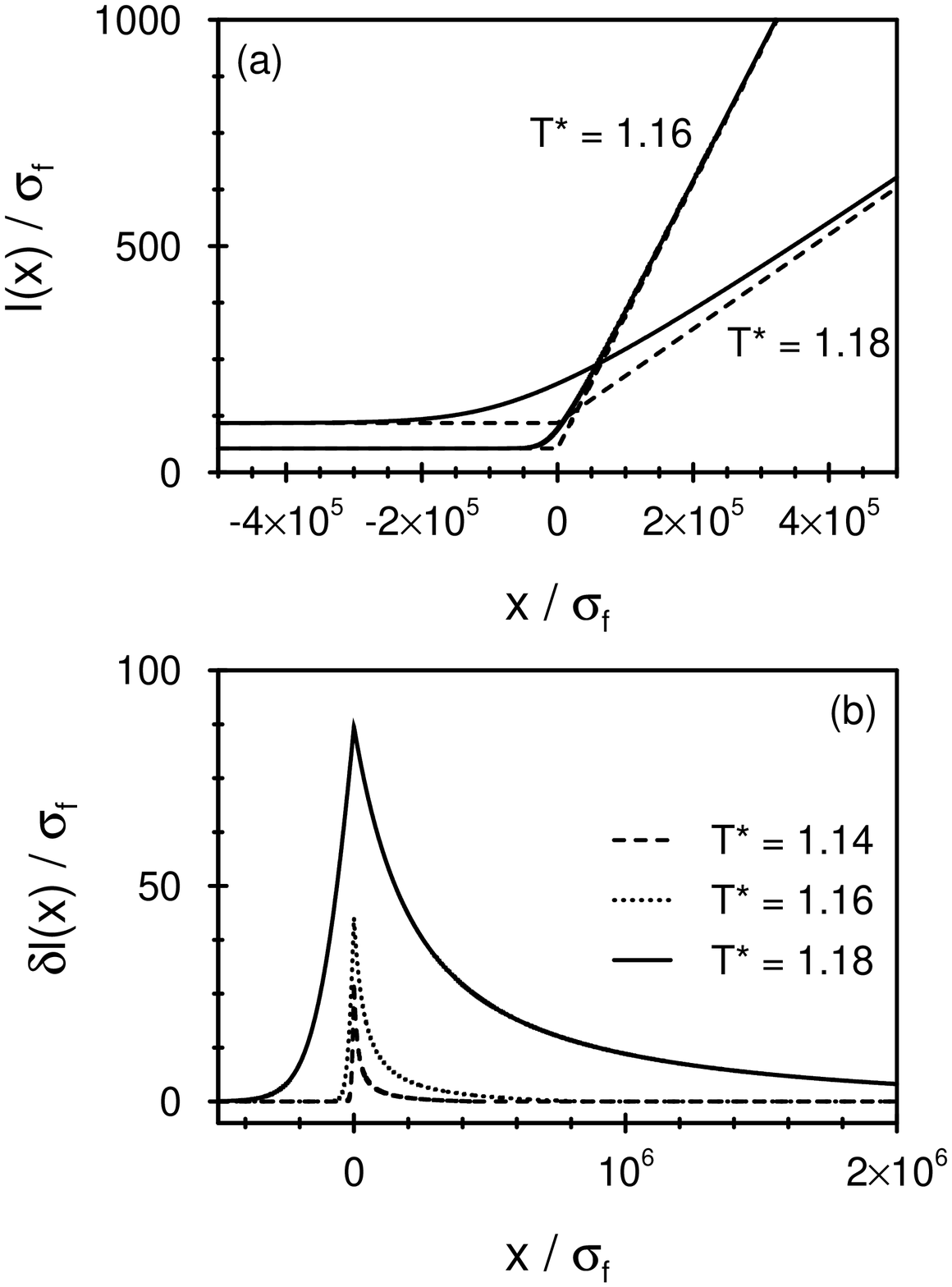, width=11cm, bbllx=65, bblly=180,
  bburx=530, bbury=815}
\epsfig{file=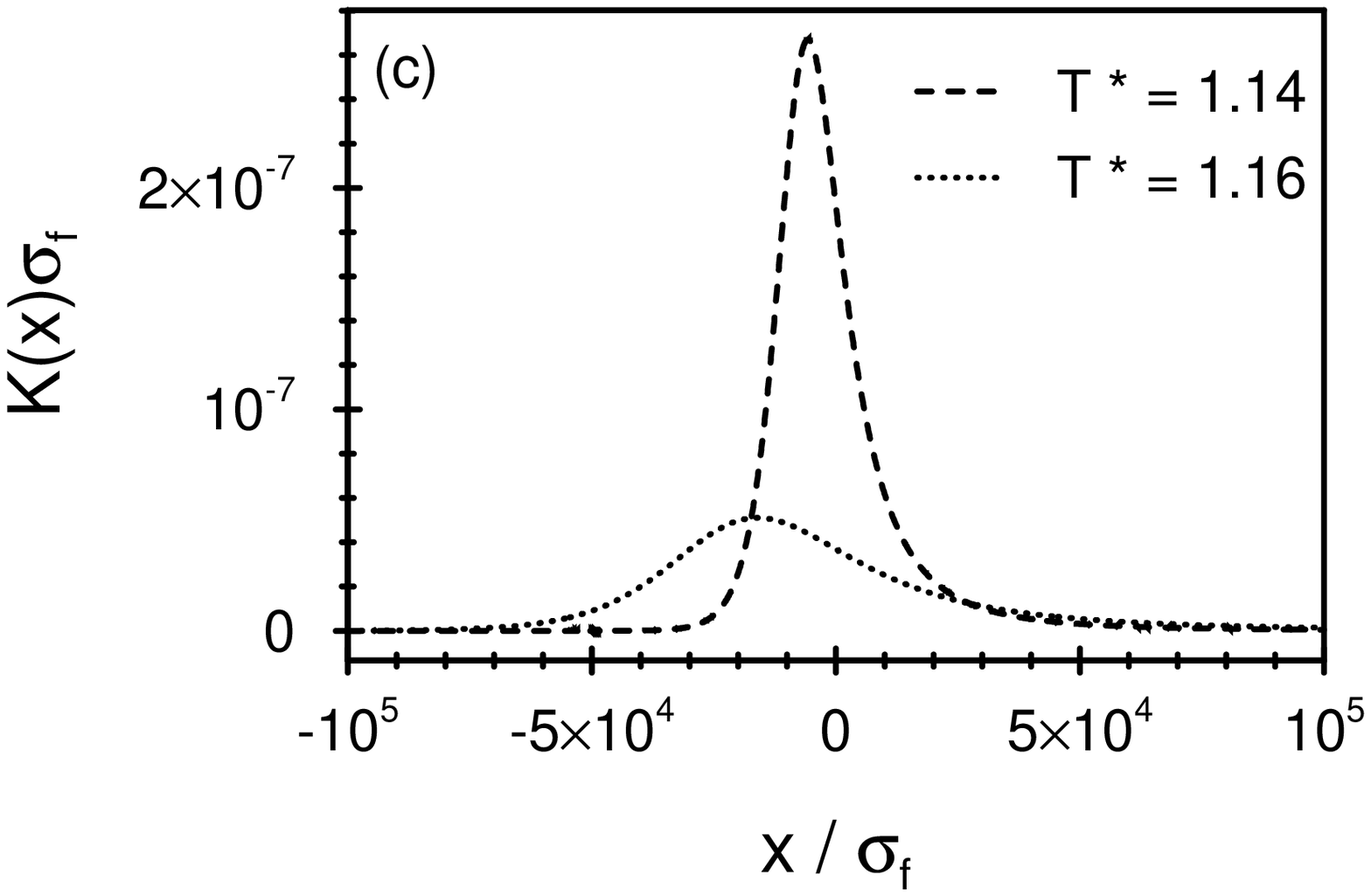, width=11cm, bbllx=65, bblly=500,
  bburx=530, bbury=830}
\end{center}
\caption{\label{f:cl_localcrit}
Liquid-vapor interfaces calculated within the local theory for a system
which exhibits a critical wetting transition at $T_w^* = 1.2$. Here
the substrate potential parameters are $d_w=\sigma_f$,
$u_3=1.683\epsilon_f\sigma_f^3$, $u_4=10.098\epsilon_f\sigma_f^4$, and 
$u_9=0.224\epsilon_f\sigma_f^9$. In (a) the interface profiles $l(x)$
(full lines) and their asymptotes $a(x)$ (dashed lines)
are shown, whereas (b) displays the deviation $\delta l(x) =
l(x)-a(x)$ of $l(x)$ from its asymptotes. The corresponding contact
angles are $\theta \approx 0.31^{\circ}$, $\theta \approx 0.17^{\circ}$,
and $\theta \approx 0.06^{\circ}$ for $T^* = 1.14$, $1.16$, and $1.18$,
respectively. $l(x)$ approaches its asymptotes from above both for
$x\to-\infty$ and $x\to\infty$. In (c) the curvatures $K(x)$ are
plotted. As compared to the system exhibiting a first-order wetting
transition (see Figs.~\ref{f:cl_local1st} --
\ref{f:cl_localnlocal_curv}) $\delta l(x)$ and the radii of curvature
$1/K(x)$ are extremely large . Due to
numerical difficulties no data for the nonlocal theory are available
for this system. However, since the interface curvature is several
orders of magnitude smaller than in the system which displays a
first-order wetting transition -- with already small deviations of the
local from the nonlocal theory -- the difference between local and
nonlocal results for the present system is expected to be indistinguishable
on this scale.} 
\end{figure}

\begin{figure}
\begin{center}
\epsfig{file=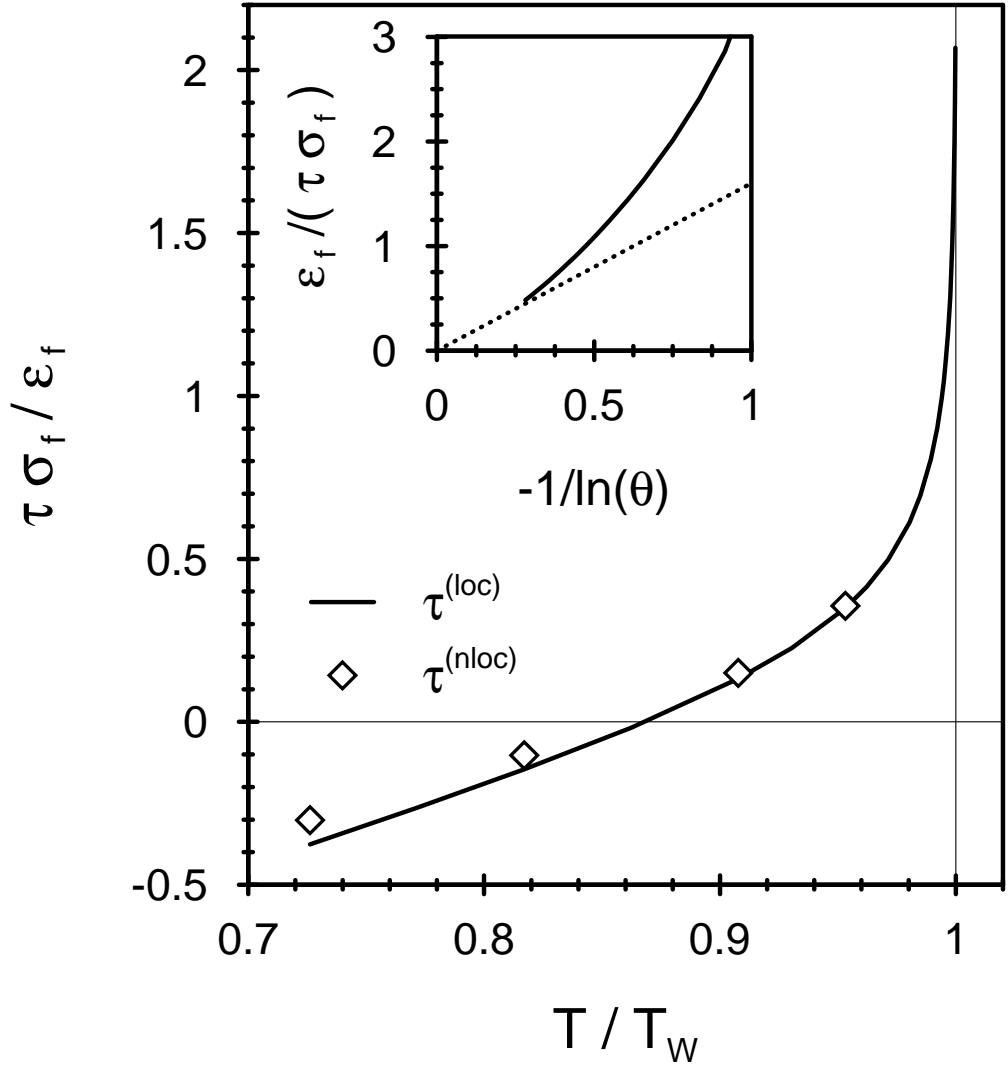, width=13cm, bbllx=80, bblly=250, bburx=520, bbury=740}
\end{center}
\caption{\label{f:cl_lt}
Line tensions as predicted by the local and the nonlocal theory for the
same system as in Fig.~\ref{f:cl_local1st}. The transition temperature
for first-order wetting is $T_w^* \approx 1.102$. The difference
$\tau^{(nloc)}-\tau^{(loc)}$ 
vanishes upon approaching $T_w$. The singular behavior of the line tension
can be derived analytically within the local theory; upon
approaching the wetting transition $\tau$ diverges as
$\tau(T \to T_w) \sim \ln(\theta(T))$, as indicated in the inset. (The
dotted line is the estimated extrapolation according to the
logarithmic divergence.) We note that as function of temperature
$\tau$ changes sign. Apart from the close vicinity of $T_w$ and its
zero the absolute value of $\tau$ is of the order of
$\epsilon_f/\sigma_f$.}
\end{figure}

\begin{figure}
\begin{center}
\epsfig{file=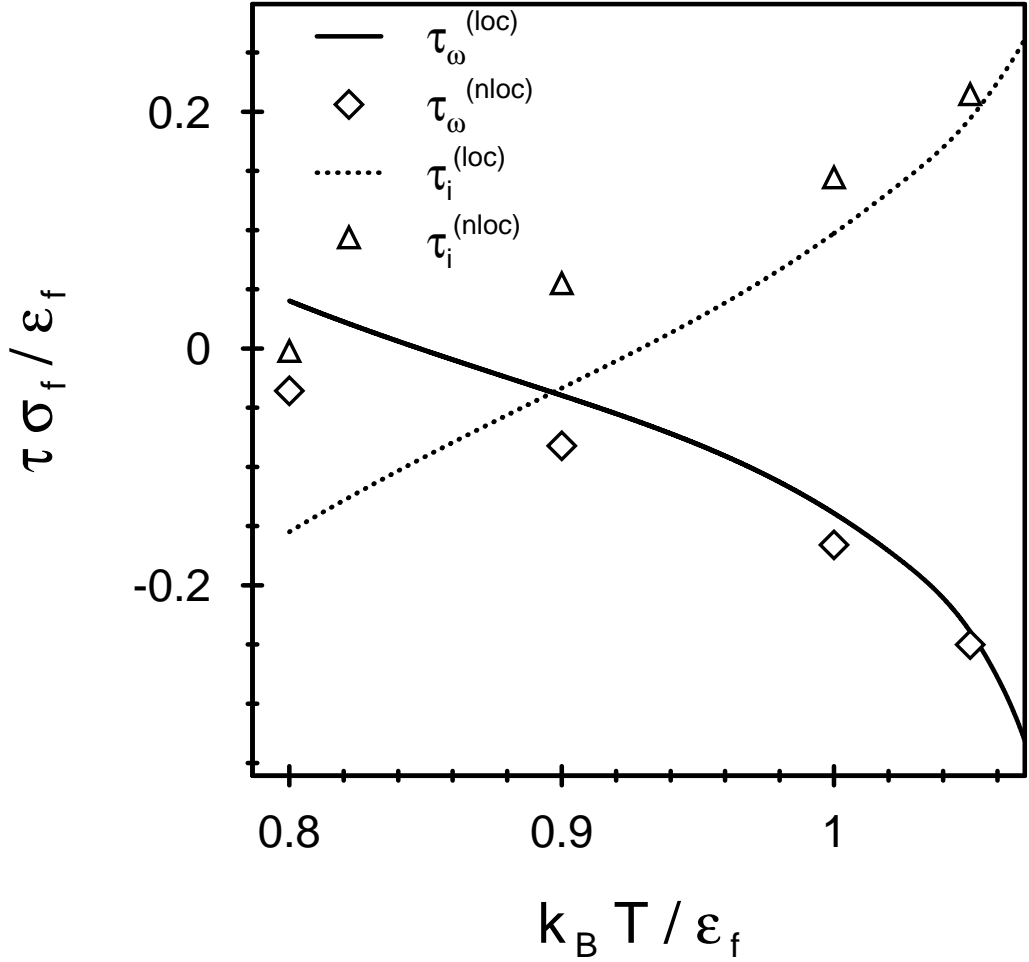, width=13cm, bbllx=85, bblly=340, bburx=515,
  bbury=770} 
\end{center}
\caption{\label{f:cl_contriblt}
Temperature dependence of the contributions $\tau_{\omega}$ and
$\tau_i$ (Eqs.~(\ref{e:cl_subdivtau})-(\ref{e:cl_locfunc})) to the
line tensions $\tau^{(nloc)}$ and $\tau^{(loc)}$ for 
the same system as in Figs.~\ref{f:cl_local1st} and \ref{f:cl_lt}. For
$T \to T_w \approx 1.102\epsilon_f/k_B$ both
$\tau_{\omega}^{(nloc)}-\tau_{\omega}^{(loc)}$ and 
$\tau_i^{(nloc)}-\tau_i^{(loc)}$ vanish. As already discussed in
Ref.~\protect\cite{getta}, the leading singular behaviors $\tau_{\omega}
\sim -1/\theta$ and $\tau_i \sim 1/\theta$ for $\theta\to 0$, i.e.,
$T\to T_w$ compensate each other in $\tau=\tau_{\omega}+\tau_i$, leaving
the residual singularity $\tau \sim -\ln(\theta)$.}
\end{figure}

\begin{figure}
\begin{center}
\epsfig{file=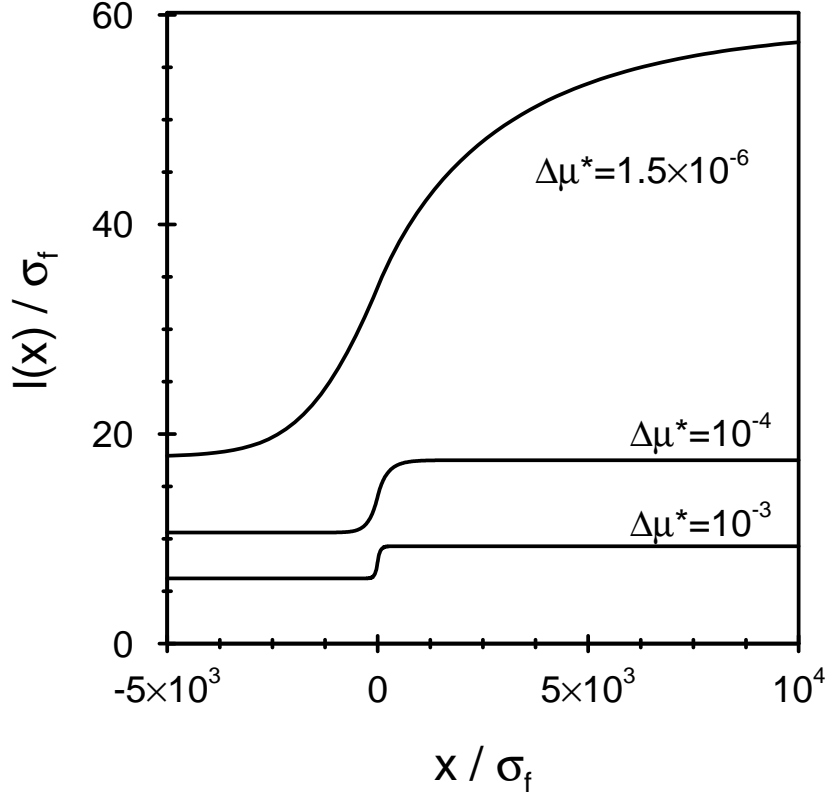, width=11cm, bbllx=95, bblly=345, bburx=540, bbury=780}
\end{center}
\caption{\label{f:step_complete}
Shape of the liquid-vapor interface on a planar substrate exhibiting a
chemical step (see Fig.~\ref{f:step_system}) within nonlocal and local
theory along an isotherm $(T^*=1.1,\Delta\mu^*=\Delta\mu/\epsilon_f\to
0)$, i.e., for complete wetting. The parameters for the
substrate potential are chosen such that both substrates individually
exhibit critical wetting transitions, the substrate $w_-$ ($x<0$) at $T_w^*
= 1.2$ and the substrate $w_+$ ($x>0$) at $T_w^* = 1.0$: 
$d_w^+=\sigma_f$, $u_3^+=2.079\epsilon_f\sigma_f^3$,
$u_{4,z}^+=u_{4,x}^+=12.475\epsilon_f\sigma_f^4$, and  
$u_9^+=0.277\epsilon_f\sigma_f^9$, whereas the parameters $d_w^-$ and
$u_j^-$ are chosen as in Fig.~\ref{f:cl_localcrit} with
$u_{4,x}^-=u_{4,z}^-$. Therefore for this temperature
at coexistence $\Delta\mu=0$ the substrate $w_+$ is wet whereas the
substrate $w_-$ remains only partially wet so that
$l(x\to\infty,\Delta\mu=0)=\infty$ and $l(x\to-\infty,\Delta\mu=0) <
\infty$. The nonlocal result for $\Delta\mu =
1.5\cdot10^{-6}$ has been obtained by solving the discretized version
of the ELE on lattices of different mesh size $\delta x$ and the application
of a pointwise extrapolation scheme $\delta x\to 0$ as outlined in the
main text. The other interfacial shapes calculated within the nonlocal theory
result from a free numerical minimization of the line contribution to
the free energy $\Omega_l^{(nloc)}[l(x)]$. The differences between
the local and the nonlocal results are not visible on this
scale. Therefore only one type of line is shown for each interfacial
profile.}  
\end{figure}

\begin{figure}
\begin{center}
\epsfig{file=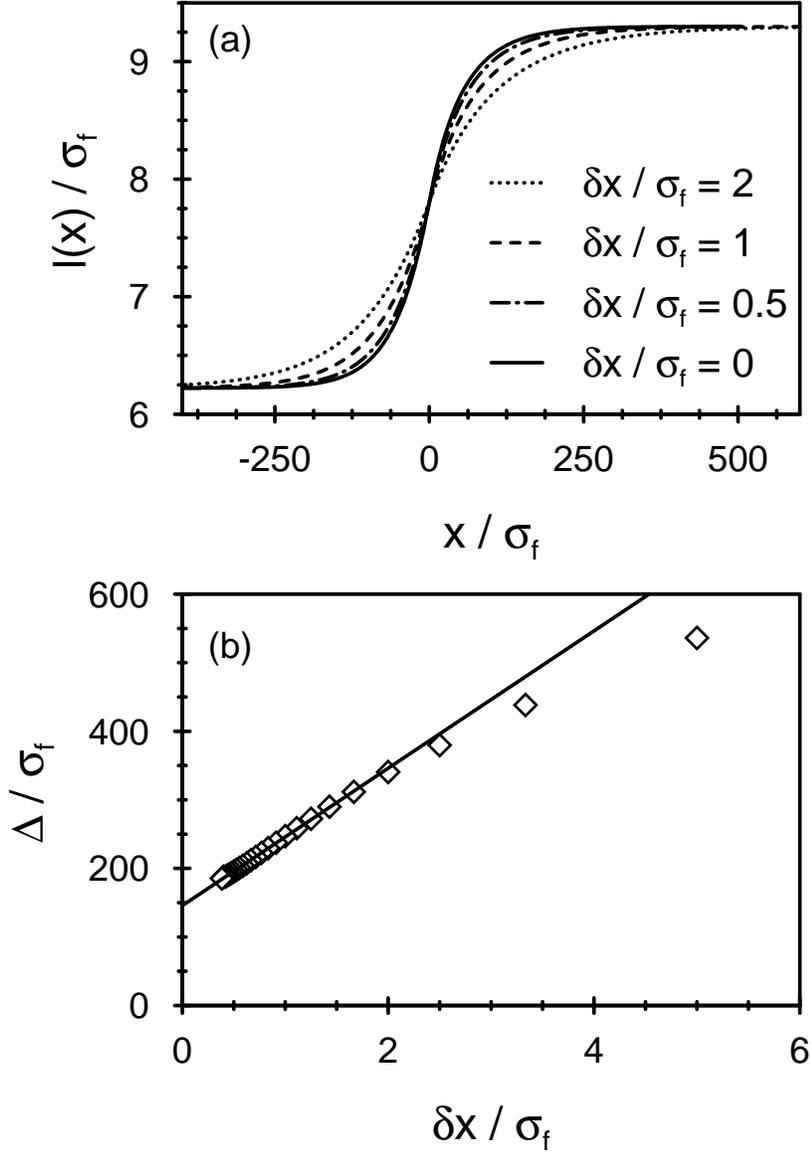, width=11cm, bbllx=85, bblly=180,
  bburx=540, bbury=825}
\end{center}
\caption{\label{f:step_ndependence}
Dependence of the numerical solutions of the nonlocal ELE
on the mesh size $\delta x$ for the same set
of parameters as in Fig.~\ref{f:step_complete} and for $\Delta\mu^* =
10^{-3}$. (a) displays profiles obtained for different values of $\delta
x$. As $\delta x$ is decreased, the distance $\Delta$ between the two
points where $l(x)$ deviates from the asymptotes $l_+$ and $l_-$ by
$10\%$ becomes smaller, as shown in (b). This dependence can be
extrapolated to the limit $\delta x = 0$, as indicated by the straight
line in (b). Using the value of $\Delta$ from this extrapolation, the
correct result in the limit of an infinitely fine lattice can be
obtained along the lines discussed in the main text. This limiting
profile is shown in (a) ($\delta x = 0$); it turns out to be
indistinguishable from both the local result and the result obtained
by the full numerical minimization of $\Omega_l^{(nloc)}$ which can also
be carried out for the present example.}
\end{figure}

\begin{figure}
\begin{center}
\epsfig{file=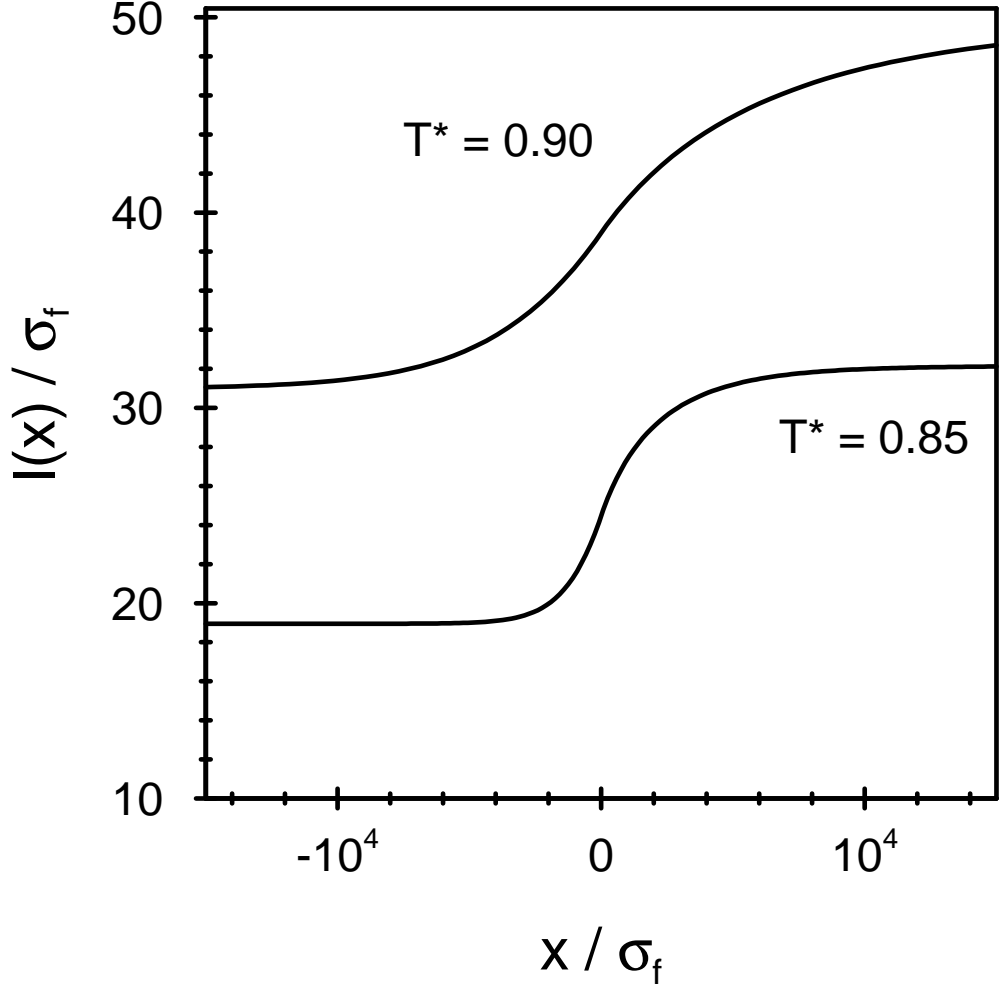, width=13cm, bbllx=95, bblly=340,
  bburx=515, bbury=770}
\end{center}
\caption{\label{f:step_critical}
Liquid-vapor interfaces on a planar substrate with a chemical step
(see Fig.~\ref{f:step_system}) as obtained within the nonlocal and the
local theory. The parameters for the
substrate potential are chosen such that both substrates
exhibit critical wetting at the same transition temperature $T_w^* =
1.0$. The parameters $d_w^+$ and $u_j^+$ are chosen as in
Fig.~\ref{f:step_complete}, whereas $d_w^-=\sigma_f$, $u_3^-=u_3^+$,
$u_9^-=u_9^+$, and $u_{4,z}^+=u_{4,x}^+=15.594\epsilon_f\sigma_f^4$. The
profiles are calculated for different temperatures $T^*$ on a
thermodynamic path along 
coexistence $\Delta\mu=0$. The nonlocal results for both temperatures
have been obtained by applying the extrapolation scheme explained in
the main text. On the present scale the local and nonlocal results are
indistinguishable and therefore only one type of lines is shown.}
\end{figure}

\begin{figure}
\begin{center}
\epsfig{file=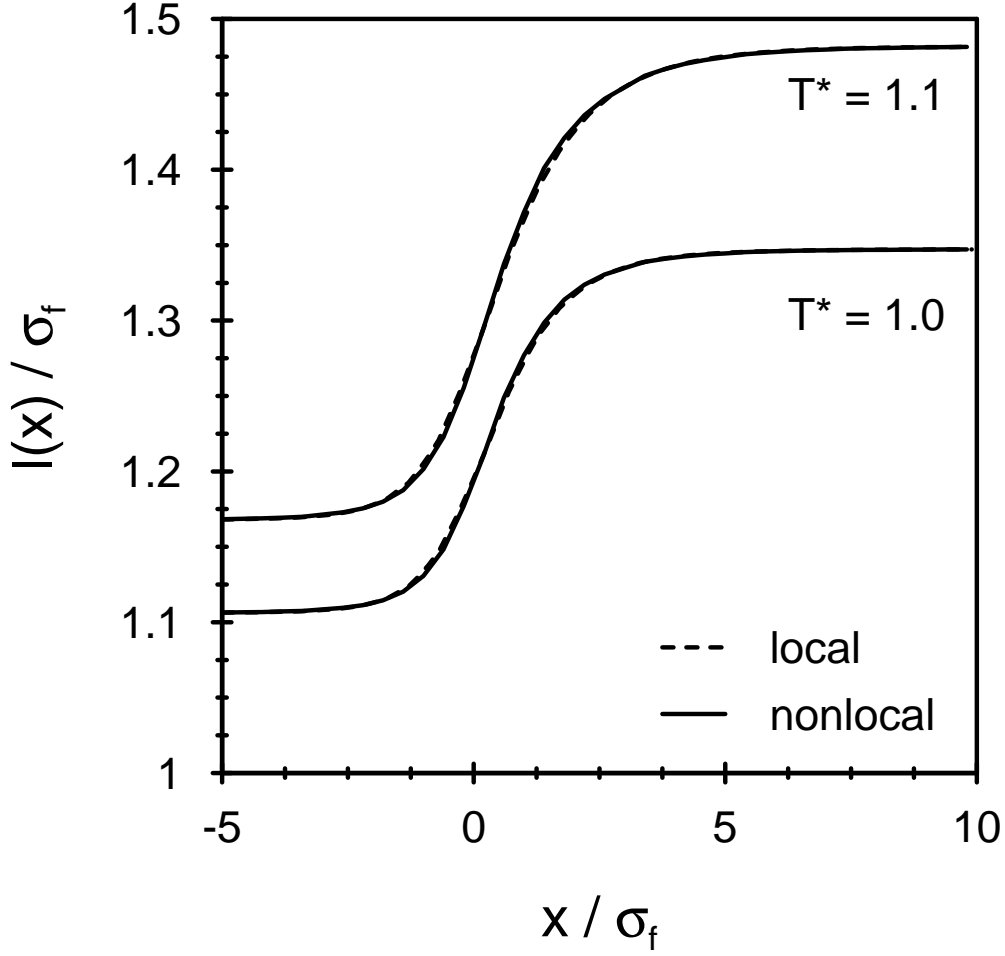, width=13cm, bbllx=85, bblly=320,
  bburx=525, bbury=775} 
\end{center}
\caption{\label{f:step_1storder}
Liquid-vapor interfaces on a substrate with a chemical step at $x=0$
(see Fig.~\ref{f:step_system}) as obtained within the nonlocal and the
local theory. The parameters for the
substrate potential are chosen such that both substrates individually
exhibit first-order wetting transitions, the substrate $w_-$ at $T_w^*
\approx 1.314$ (with $d_w^-=\sigma_f$, $u_3^-=2.513\epsilon_f\sigma_f^3$,
$u_{4,z}^-=u_{4,x}^-=3.770\epsilon_f\sigma_f^4$, and  
$u_9^-=0.335\epsilon_f\sigma_f^9$) and the substrate $w_+$ at $T_w^* \approx
1.102$ (with the parameters $d_w^+$ and $u_j^+$ as for
Fig.~\ref{f:cl_local1st} and $u_{4,x}^+=u_{4,z}^+$). The
system is at coexistence $\Delta\mu=0$. 
Within the nonlocal and the local theory both profiles are obtained by a full
numerical minimization of the line contribution to the free energy
$\Omega_l^{((n)loc)}$. The differences between the local 
and the nonlocal results are very small but detectable. The relative
difference in line tensions
$(\tau^{(nloc)}-\tau^{(loc)})/\tau^{(loc)}$ is of the order of $10^{-2}$.}
\end{figure}

We have analyzed the morphology and the associated line tensions of
liquidlike wetting films at a three-phase contact line
(Fig.~\ref{f:cl_system}) and on a planar substrate across a chemical
step (Fig.~\ref{f:step_system}). By using refined numerical techniques
we have compared quantitatively the predictions obtained within a
local displacement model for the interface profile and within a
nonlocal density functional theory. Based on general arguments
\cite{napdiet1,napdiet2,napdiet3,napdiet4} the latter approach is the
more accurate one. We have obtained the following main results:
\begin{enumerate}
\item Within the present mean-field theories the equilibrium
  interfacial profiles are determined as the minimum of the line
  contribution to the functional of the grand canonical free
  energy. These equilibrium profiles, within the sharp-kink
  approximation (Fig.~\ref{f:cl_sharpkink}), can be obtained either by
  an explicit minimization procedure or by solving the
  corresponding Euler-Lagrange equation for the minimum. Within the
  local theory both approaches are equally successful and
  robust. However, within the nonlocal theory in practical terms only
  the minimization procedure yields reliable access to the true
  minimum whereas solving the necessarily discretized version of the
  nonlocal Euler-Lagrange equation leaves one with a very slow
  convergence with respect to the mesh size of the
  discretization. This mesh size dependence of the solutions within the nonlocal
  theory was not revealed in Refs.~\cite{getta} and \cite{koch} and
  led to a significant overestimation of the quantitative discrepancy
  between the predictions of the local and the nonlocal theory.
\item Concerning the intrinsic structure of the three-phase contact
  line the local and the nonlocal theory yield only slightly different
  interface morphologies (Figs.~\ref{f:cl_local1st} and
  \ref{f:cl_localnlocal1st}) and line tensions (Figs.~\ref{f:cl_lt} and
  \ref{f:cl_contriblt}) in the case of first-order wetting. In the
  case of critical wetting the two theories yield indistinguishable
  results (Fig.~\ref{f:cl_localcrit}). For first-order wetting the
  deviations between the results obtained from the local and the
  nonlocal theories are confined to regions within which the radius of
  curvature of the interface profile is less than about 10 atomic
  diameters $\sigma_f$ of the fluid particles
  (Fig.~\ref{f:cl_localnlocal_curv}). Upon approaching a first-order
  wetting transition the line tension $\tau$ of the three-phase
  contact line diverges logarithmically (Figs.~\ref{f:cl_lt} and
  \ref{f:cl_contriblt}). $\tau$ changes sign as a function of
  temperature and typically is of the order of $\epsilon_f/\sigma_f$
  where $\epsilon_f$ is the depth of the interaction potential between
  the fluid particles.
\item Concerning the morphology of a wetting film on a planar
  substrate with a chemical step the results obtained from the local
  and the nonlocal theory for complete and critical wetting are
  indistinguishable (Figs.~\ref{f:step_complete} and
  \ref{f:step_critical}). Small differences appear only in the case of
  first-order wetting (Fig.~\ref{f:step_1storder}).
\end{enumerate}
Thus we conclude that the local interface displacement model yields
quantitatively reliable results for the interface morphology and the
line tension of laterally inhomogeneous wetting films. The more
accurate nonlocal description is necessary only for cases in which the
interface profiles exhibit large curvatures. Thus all results and
discussions in Refs.~\cite{getta} and \cite{koch} based on the local
theory turn out to provide actually a quantitatively reliable
description of the systems under consideration.

\acknowledgements

We gratefully acknowledge many useful discussions with T.~Boigs concerning the
reliability of the numerical solutions of the nonlocal Euler-Lagrange
equations.

\pagebreak

\end{document}